\documentclass[a4paper, amsfonts, amssymb, amsmath, reprint, showkeys, nofootinbib, twoside,prx]{revtex4-2}
\usepackage[english]{babel}
\usepackage[utf8]{inputenc}
\usepackage[colorinlistoftodos, color=green!40, prependcaption]{todonotes}
\usepackage[pdftex, pdftitle={Article}, pdfauthor={Author}]{hyperref} 
\usepackage{tabularx}
\begin{document}
\title{Tricritical Dicke model with and without dissipation}

\author{Diego Fallas Padilla}
\email{daf5@rice.edu}
\affiliation{Department of Physics and Astronomy, and Rice Center for
Quantum Materials, Rice University, Houston, Texas 77251-1892, USA} 
\author{Han Pu}
\email{hpu@rice.edu}
\affiliation{Department of Physics and Astronomy, and Rice Center for
Quantum Materials, Rice University, Houston, Texas 77251-1892, USA}

\begin{abstract}
Light-matter interacting systems involving multi-level atoms are appealing platforms for testing equilibrium and dynamical phenomena. Here, we explore a tricritical Dicke model, where an ensemble of three-level systems interacts with a single light mode, through two different approaches: a generalized Holstein-Primakoff mapping, and a treatment using the Gell-Mann matrices. Both methods are found to be equivalent in the thermodynamic limit. In equilibrium, the system exhibits a rich phase diagram where both continuous and discrete symmetries can be spontaneously broken. We characterize all the different types of symmetries according to their scaling behaviors. Far from the thermodynamic limit, considering just a few tens of atoms,  the system already exhibits features that could help characterize both second and first-order transitions in a potential experiment. Importantly, we show that the tricritical behavior is preserved when dissipation is taken into account. Moreover, the system develops a rich steady-state phase diagram with various regions of bistability, all of them  converging at the tricritical point. Having multiple stable normal and superradiant phases opens prospective avenues for engineering interesting steady states by a proper choice of initial states and/or parameter quenching. 
\end{abstract}

\date{\today }
\maketitle

\section{Introduction}

When $N$ two-level atoms confined in a small volume interacting with a single mode of light are all initialized in their excited state, their spontaneous emission processes can interfere constructively leading to an intensity-enhanced pulse of emitted light. This concept, known as Dicke superradiance, was introduced in 1954 \cite{Dicke1954} and represents a foundation for numerous subsequent studies regarding the coherence between emitters across several platforms \cite{gross1976,scheibner2007superradiance,skribanowitz}.

Later, an equilibrium notion of superradiance was introduced with the Dicke model \cite{hepplieb,hioe}. In the limit of an infinite number of atoms, and above a critical value of the light-matter interaction strength, this model undergoes a second-order phase transition from a normal phase with a vanishing photon population to a superradiant phase with a macroscopic photon population. Near the critical point, interesting features such as squeezing \cite{masson2019,hayashida2023perfect} or the onset of chaotic behavior \cite{Emary2003} are expected. 

Several extensions of the Dicke model have been proposed to unlock new exotic phenomena, of particular interest is the generalization to multi-level atoms \cite{Xu1,skulte,haynthermo,fan2023}. Having more than two atomic levels allows for creative model proposals where the connectivity between levels can be engineered to generate valuable properties, such as the generation of dark states \cite{Hayn2011} or the unlocking of different dynamical phases \cite{Valencia2023}. In this work, we focus on the study of multicritical points, specifically, tricritical points (TPs), using multi-level Dicke models \cite{Youjiang2021}. A TP signals the intersection of a first- and a second-order phase transition, and with these two types of transitions having very different behaviors, a highly tunable system exhibiting a TP is ideal for exploring universal scaling, hysteresis, and metastability that goes beyond the much more extensively explored second-order quantum criticality. Recently, evidences of TPs have been reported in magnetic materials \cite{kaluarachchi,Kim2002,kashanov2018,QTPex}. Such systems, however, lack the parameter tunability available in atomic/optical platforms. 

Realizing models in a cavity QED environment typically requires an open system description due to the necessary presence of one or more sources of losses. Open Dicke-like systems are of great interest in their own right. For example, they show interesting steady-state features not present in the closed systems \cite{nagy2011,Bhaasen,gelhausen,Soriente,Gegg_2018} as the inclusion of dissipation channels can modify the geometry of the phase boundaries, alter the order of the transitions, and generate regions of multi-stability where the final state of the system is highly dependent on initial state preparation.  

In this work, we introduce a tricritical Dicke model (TDM), describing a single cavity mode interacting with an ensemble of three-level atoms. In this model, both continuous and discrete symmetry breakings can occur. First, we characterize the equilibrium phase diagram and critical scaling in the thermodynamic limit. Second, we explore the system away from the thermodynamic limit with finite number of atoms. Lastly, we describe the non-equilibrium phase transition landscape in the presence of losses and examine how the TP manifests in such an open system. The richness of the dissipative phase diagram characterized by different types of phase transitions and regions of bistability opens exciting possibilities for engineering desired steady states.

\section{Model} 
\begin{figure}[t!]
\includegraphics[width=0.48\textwidth]{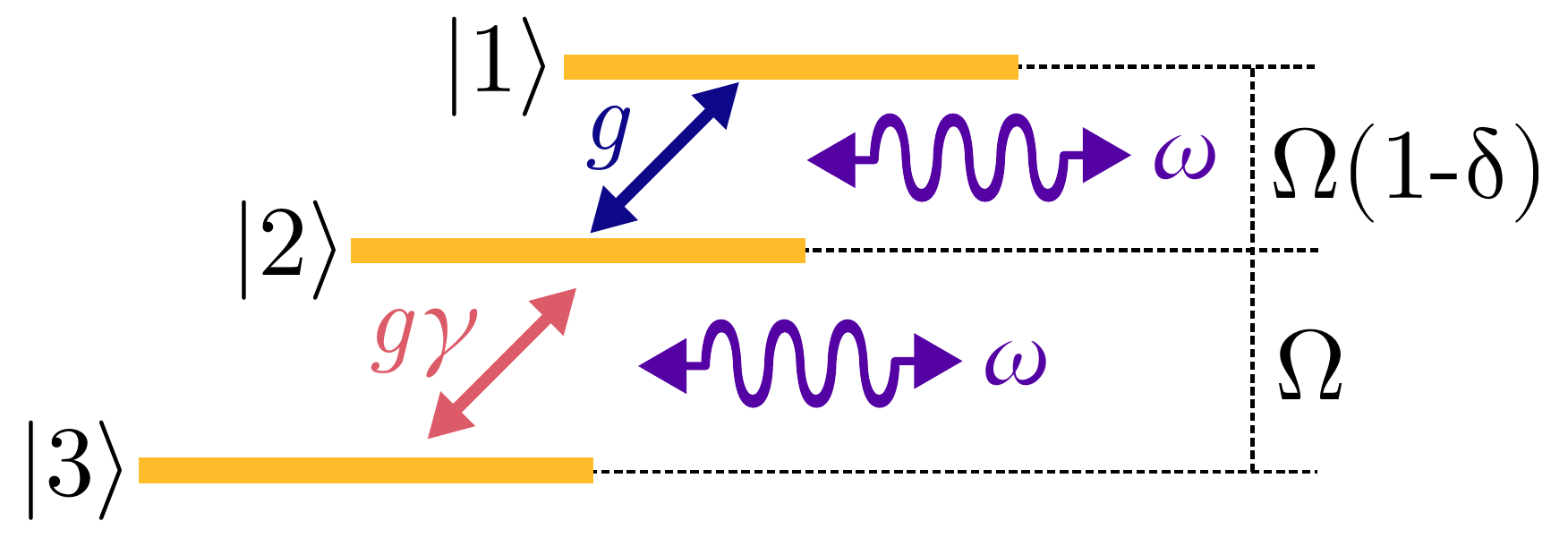}
\caption{Schematics of the TDM. The three states $\vert 1 \rangle$, $\vert 2 \rangle$, and $\vert 3 \rangle$ are represented by the top, middle, and bottom yellow horizontal bars, respectively. Wavy arrows represent photons of frequency $\omega$. The light-matter interaction terms are represented by solid arrows, here,  $g=g_1$ for co-rotating terms and $g=g_2$ for counter-rotating terms. }
\label{Fig1}
\end{figure}

We consider an ensemble of $N$ three-level atoms interacting with a single mode of light. We denote the $j$-th atom level by $ \vert m \rangle^{(j)}$, with $m=1,2,3$ corresponding to spin projection values in the $z$-direction 1, 0, and -1, respectively. The TDM is a generalization of the conventional Dicke model where now the phase transition between the superradiant and normal phases can occur across second-order line, a first-order line, or a TP. The TDM is described by the Hamiltonian:
\begin{eqnarray}
    H &=& \omega a^{\dagger}a + \Omega(1-\delta)P_{11} - \Omega P_{33} \nonumber \\
    &&+\frac{g_1}{\sqrt{N}}(a (P_{12}+\gamma  P_{23})+a^{\dagger}(P_{21}+\gamma P_{32})) \nonumber \\
    &&+\frac{g_2}{\sqrt{N}}(a^{\dagger} (P_{12}+\gamma  P_{23})+a(P_{21}+\gamma P_{32})) \,.
    \label{TDMH}
\end{eqnarray}
Here $P_{mm'}$ denotes a collective atomic operator defined as $P_{mm'}=\sum_{j=1}^N \vert m \rangle^{(j)}\langle m' \vert ^{(j)}$,  $a$ ($a^{\dagger}$) denote bosonic annihilation (creation) operators for the cavity photon mode, $\omega$ is the photon frequency, $\Omega$ characterizes the atomic energy splitting,
$g_1$ is the light-matter interaction strength for the co-rotating terms and $g_2$ for the counter-rotating ones. Finally, the two dimensionless quantities $\gamma$ and $\delta$ represent an imbalance in the light-matter coupling strength and energy splitting between different atomic levels, respectively (see Fig.~\ref{Fig1}). 

Dicke-like models where co- and counter-rotating terms are allowed to have different coupling strengths have been widely explored~\cite{Soriente,DEAGUIAR1992291,FURUYA1992313,Bastarrachea-Magnani_2016,KLOC201785,Baksic}. For the present case, if $g_1=g_2$ the system reduces to a specific case of the multicritical Dicke model presented in \cite{Youjiang2021}, on the other hand, if $g_2=0$, the system reduces to a previously studied Tavis-Cummings model that also exhibits a tricritical point \cite{cordero2013}. 

For the cavity system, one obstacle on the way is that experimental observation of the superradiant phase transition in its ``pure" form is seriously challenged by the no-go theorem stated by Rza\ifmmode \dot{z}\else \.{z}\fi{}ewski {\it et al.} \cite{rzazewski}, where the inclusion of the $A^2$ term from the dipole interaction prevents the transition from occurring. One way to circumvent this no-go theorem is to consider a system where the coupling between atomic levels is achieved by cavity-assisted Raman transitions. This could be realized, for example, in an optical cavity QED system through the coupling of different atomic hyperfine magnetic sub-levels with additional lasers \cite{Dimer}. This is the scheme we adopt here.

Finally, we want to remark that several other groups have previously studied three-level atoms interacting with cavity fields~\cite{Hayn2011,leonard,Lopez2021,Cordero_2021}, see Ref.~\cite{Larson_2021} for a review. However, in those works, different atomic transitions are coupled to different cavity modes. As such, the physics exhibited in those systems are qualitatively different from ours. 

\section{Thermodynamic Limit} 

Let us first explore the thermodynamic limit in which the atom number $N \longrightarrow \infty$, while the coupling strengths $g_1$ and $g_2$ are finite. 

\subsection{Generalized Holstein-Primakoff mapping} In the conventional Dicke model, a Holstein-Primakoff mapping \cite{HP1940}, where the spin collective operators are mapped into a single bosonic mode, is often used to explore the mean-field properties of the system~\cite{Emary2003}. This mapping can be intuitively understood as promoting one two-level atom from the ground state to the excited state is equivalent to adding one quantum of excitation in the mapped bosonic mode. In the TDM, since we are dealing with three-level atoms, we require to map the atomic collective operators into two different bosonic modes through a generalized Holstein-Primakoff mapping as suggested in Ref.~\cite{Hayn2011}. In order to conduct the generalized mapping, we choose state $\vert 3 \rangle$ as our reference state, the mapping is then defined by:
\begin{eqnarray}
P_{mm'} &=& b^{\dagger}_m b_m', \quad  m,m' =1,2 \,, \nonumber \\
P_{m3} &=& b^{\dagger}_j \,\Theta = \left(P_{3m} \right)^\dag, \quad m =1,2 \,, \nonumber \\
P_{33} &=& N - \sum_{m=1,2} b^{\dagger}_m b_m \,.
\end{eqnarray}
where $\Theta \equiv \sqrt{N - b_1^{\dagger}b_1 - b_2^{\dagger}b_2}$. 
The Hamiltonian is now given by:
\begin{eqnarray}
H &=& \omega a^{\dagger}a - N \Omega + (2-\delta)\Omega \,b_1^{\dagger}b_1 + \Omega \,b_2^{\dagger}b_2 \nonumber \\
&+&\frac{g_1 }{\sqrt{N}}(ab_1^{\dagger}b_2 + a^{\dagger}b_2^{\dagger}b_1) + \frac{g_1 \gamma}{\sqrt{N}}(ab_2^{\dagger} \Theta+ a^{\dagger} \Theta b_2) \nonumber \\
&+&\frac{g_2 }{\sqrt{N}}(ab_2^{\dagger}b_1 + a^{\dagger}b_1^{\dagger}b_2) + \frac{g_2\gamma}{\sqrt{N}}(a^{\dagger}b_2^{\dagger} \Theta+ a \Theta b_2) \,,
\label{HHP}
\end{eqnarray}
The form of Eq.~(\ref{HHP}) makes evident the meaning of the new bosonic operators $b_1$ and $b_2$: they take us from the reference state to the other two states, and back. Creating an excitation in state $\vert 1 \rangle$ requires an energy $(2-\delta)\Omega$ which is the detuning with respect to the reference state $\vert 3 \rangle$, a similar argument follows for state $\vert 2 \rangle$. Moreover, note that since states $\vert 3 \rangle$ and $\vert 1 \rangle$ are not directly coupled in our Hamiltonian, ``cycle" terms such as $b_1^{\dagger}b_2$ are needed using this formalism.

\subsection{Ground state phase diagram}

\begin{figure*}[t!]
\includegraphics[width=0.95\textwidth]{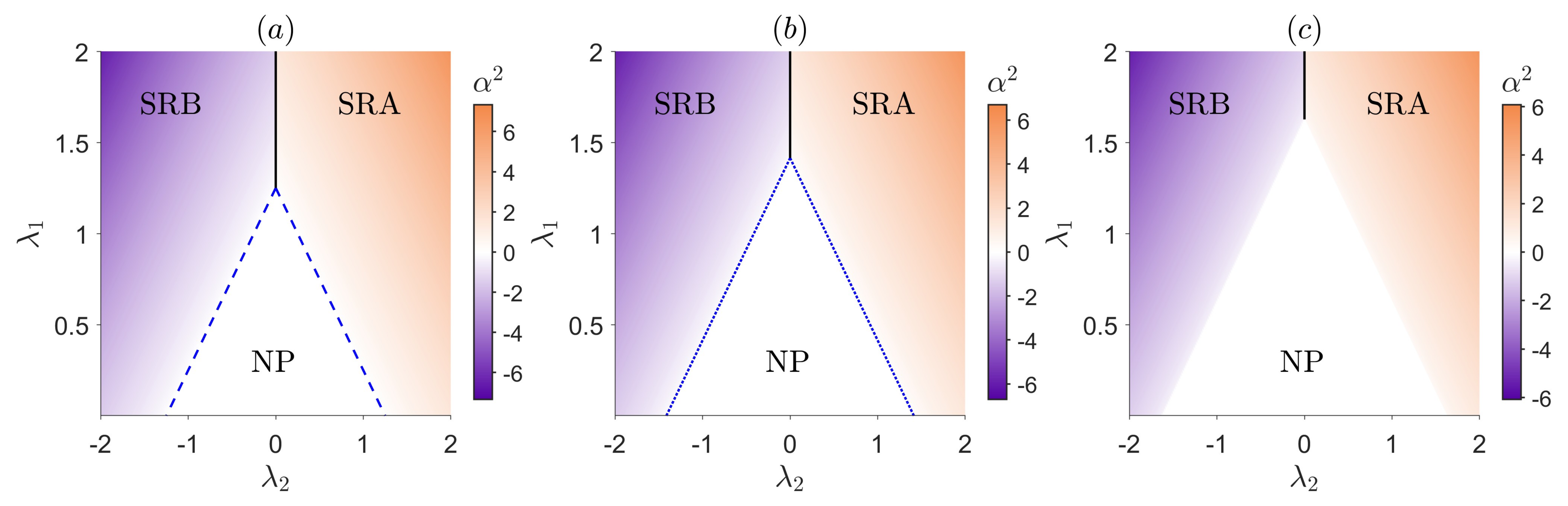}
\caption{Phase diagram in the $\lambda_1$-$\lambda_2$ plane for $\delta=0$ and (a) $\gamma =0.8$, (b) $\gamma=\gamma_{\rm TP}=1/\sqrt{2}$, and (c) $\gamma=0.6$. Here we choose $\alpha^2$ as the order parameter but equivalent phase diagrams can be constructed for $\beta_1$ and $\beta_2$. The solid vertical line represents the Tavis-Cummings line dividing the SRA and SRB phases. The dashed lines in (a) signal the second-order boundary, while the dotted lines in (b) denote the line of tricritical points, both of these lines' equations are given by Eqs.~(\ref{TPcond1}) and (\ref{TPcond2}).}
\label{Fig2}
\end{figure*}

Now, we displace each bosonic operator by their mean-field values
\begin{equation}
a = \sqrt{N} \alpha + c, \quad b_1 = \sqrt{N}  \beta_1 + d_1, \quad b_2 = \sqrt{N} \beta_2 + d_2\,,
    \label{displaced}
\end{equation}
where the mean-field values $\alpha$, $\beta_1$, and $\beta_2$ are taken to be of order $N^0$ and complex numbers, in general. The new bosonic operators $c$, $d_1$, and $d_2$ represent the variations with respect to the mean-field values. After substituting Eq.~(\ref{displaced}) into Eq.~(\ref{HHP}) and expanding in powers of $N$, the Hamiltonian can be rewritten as (see details in Appendix~\ref{AppA})
\begin{equation}
H \approx N H_0 + \sqrt{N}H_1 + H_2 \,,
\label{Eqthreeterms}
\end{equation}
where terms with negative powers of $N$ are discarded since we are considering the thermodynamic limit $N \rightarrow \infty$. The first term $H_0$ describes the ground state mean-field energy of the system and is given explicitly by:
\begin{eqnarray}
H_0 = \omega \alpha \alpha^* - \Omega + (2-\delta) \Omega \beta_1 \beta_1^* + \Omega \beta_2 \beta_2^* \nonumber \\
+ g_1(\alpha \beta_1^* \beta_2 + \textnormal{c.c}) + g_2(\alpha \beta_2^* \beta_1 + \textnormal{c.c}) \nonumber \\
+ g_1 \gamma \beta (\alpha \beta_2^* + \textnormal{c.c}) + g_2 \gamma \beta (\alpha \beta_2 + \textnormal{c.c}) \,,
\label{H0}
\end{eqnarray}
where $\beta\equiv \sqrt{1-\vert \beta_1 \vert ^2 - \vert \beta_2 \vert^2}$. Minimization of $H_0$ with respect to the real and imaginary parts of $\alpha$, $\beta_1$, and $\beta_2$ can be performed to determine the values of these parameters. The normal phase (NP) characterized by $\alpha=\beta_1=\beta_2=0$, namely, all atoms in state $\vert 3 \rangle$ and zero photon population, is always a solution to the set of equations $\partial H_0 / \partial \mu =0$ with $\mu = \alpha, \,\beta_1, \,\beta_2$. However, this phase does not always represent the configuration that minimizes the energy, in that case, the equilibrium phase becomes a superradiant phase with nonzero values of the three order parameters $\alpha$, $\beta_1$, and $\beta_2$.

For $g_1,\,g_2 \neq 0$ two different superradiant phases are found. When $g_1$ and $g_2$ have the same sign, we find that all three order parameters are real, and we denote this phase as superradiant phase A (SRA). On the other hand, when $g_1$ and $g_2$ have opposite signs, we find that $\beta_1$ remains real but both $\alpha$ and $\beta_2$ become purely imaginary, and we denote this phase as superradiant phase B (SRB). A similar behavior of  order parameters was also found in previous studies in a model interpolating between the conventional Dicke and Tavis-Cummings models \cite{Baksic,Soriente}.

For these two superradiant phases, $\alpha$ is described by a single real number and then it is possible to find the location of the TP and the equation for the second-order line by doing a single parameter Landau theory analysis after performing time-independent perturbation theory following a procedure similar to the one described in Ref.~\cite{Youjiang2021} (see Appendix~\ref{AppB}). The critical line between the normal phase and each of the superradiant phases is determined by two constraints:
\begin{equation}
\text{SRA:}\quad \gamma^2 = \frac{1}{\lambda_+^2}\geq \frac{1}{2-\delta} \,, 
\label{TPcond1}
\end{equation}
\begin{equation}
\text{SRB:}\quad\gamma^2 = \frac{1}{\lambda_-^2}\geq \frac{1}{2-\delta} \,,
\label{TPcond2}
\end{equation}
where $\lambda_\pm \equiv \vert \lambda_1 \pm \lambda_2 \vert$, and $\lambda_i \equiv g_i /\sqrt{\omega \Omega}$ are renormalized dimensionless coupling strengths. The location of the TP is obtained when the equal signs are taken in the above. Clearly, if $\delta$ is positive we require $\delta < 2$ for all parameters to be kept real. Moreover, the derivation of Eqs.~(\ref{TPcond1}) and (\ref{TPcond2}) assumes non-degenerate perturbation theory requiring $\delta \neq 1$. To reduce the number of parameters and facilitate the visualization of the different phase boundaries, we constrain ourselves to $\delta=0$. However, we will keep $\delta$ in all our derivations since in certain experimental setups it might be easier to vary this detuning instead of the parameter $\gamma$.

In Fig.~\ref{Fig2}, the phase diagram for $\delta=0$ and three different values of $\gamma$ is presented. Note that we have chosen $\alpha^2$ instead of $\vert \alpha \vert^2$ as the order parameter in order to differentiate between the SRA and SRB. For $\delta=0$ the TP is located at $\gamma = \gamma_{\rm TP}=1/\sqrt{2}$ as deducted from Eqs.~(\ref{TPcond1}) and (\ref{TPcond2}). Panel (a) illustrates how for values of $\gamma > \gamma_{\rm TP}$ the phase transition is of second order with the phase boundary defined by $\gamma^2 = 1/\lambda_{\pm}^2$. In panel (b) the transition between the NP and SRA/SRB is given by a line of TP's. Finally, in panel (c) the transition is found to be of first order as the order parameter changes discontinuously to a nonzero value across the phase transition. The phase diagram looks almost identical when $\delta \neq 0$ with the only difference being the value of $\gamma$ where the order of the phase transition changes ($\gamma_{TP}$). If $\delta$ decreases the value of $\gamma_{TP}$ decreases and vice-versa, as described in Eqs.~(\ref{TPcond1}) and (\ref{TPcond2}).

The phase diagram showcases both discrete and continuous symmetry breaking. First, note that the energy in Eq.~(\ref{H0}) is invariant under the transformation $\alpha \rightarrow - \alpha$, $\beta_2 \rightarrow -\beta_2$, $\beta_1\rightarrow \beta_1$. This $Z_2$ symmetry is spontaneously broken in the SRA/SRB phases. On the other hand, when $g_2=0$ the system is reduced to a tricritical Tavis-Cummings model, in which case $H_0$ is invariant under a more general transformation $\alpha \rightarrow \alpha e^{i\theta}$, $\beta_2 \rightarrow \beta_2 e^{i \theta}$, $\beta_1 \rightarrow \beta_1 e^{2 i \theta}$, with $\theta \in [0, 2 \pi)$. This means that there are infinite equilibrium configurations with the three order parameters being nonzero for $\lambda_+ =\lambda_- > \lambda_c$, with $\lambda_c$ being the value at which the first-order, second-order, or tricritical phase transition occurs. These solutions spontaneously break the continuous $U(1)$ symmetry. The special case $g_1=0$ is equivalent to the Tavis-Cummings case described above after a rotation of the atomic spin operators is performed \cite{kirton2019}.

\subsection{Critical behavior}

\begin{figure}[t!]
\includegraphics[width=0.48\textwidth]{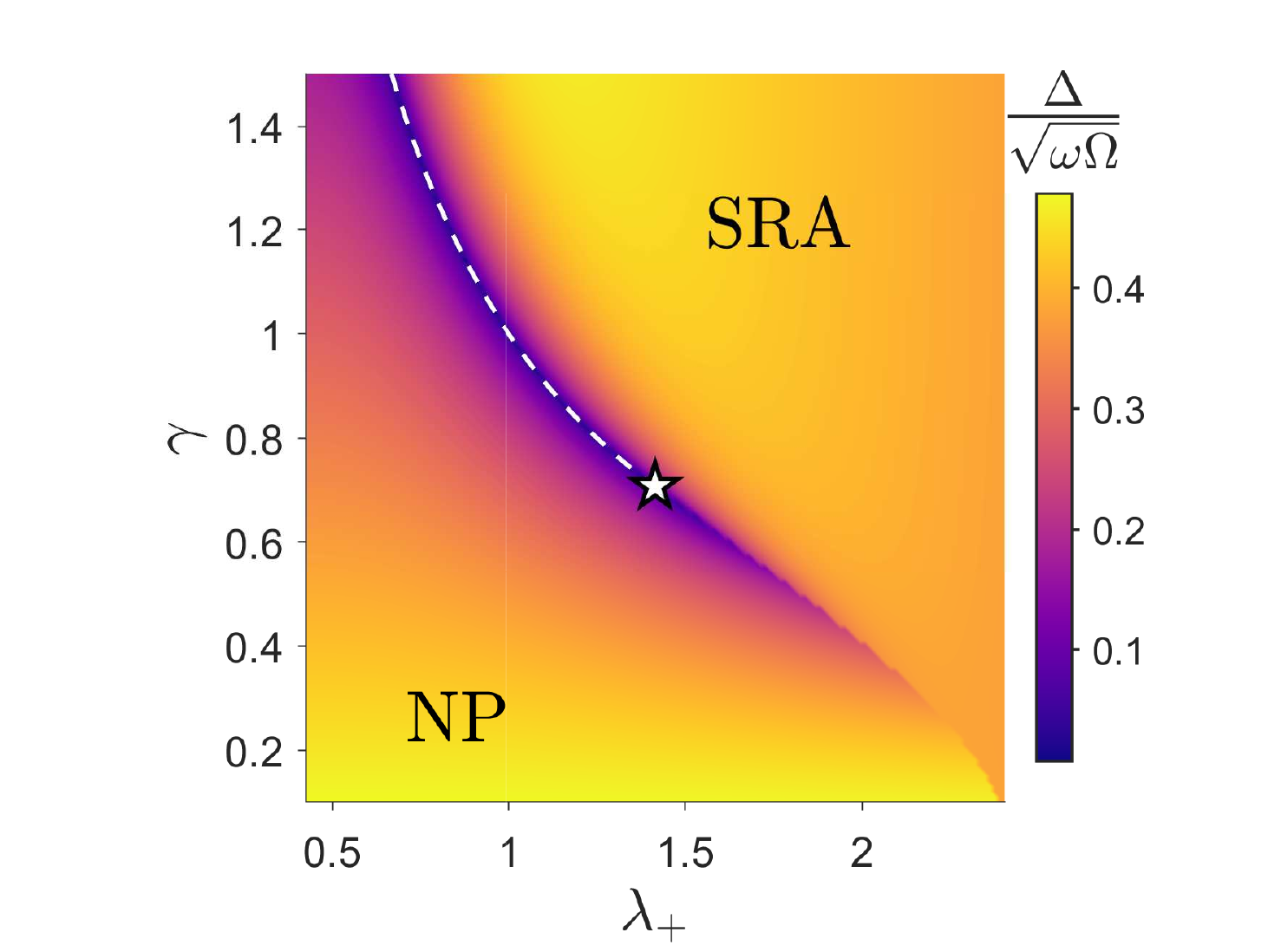}
\caption{Energy gap $\Delta$ between the ground state and first excited state in the $\gamma$-$\lambda_+$ plane for $\delta=0$ and fixed $\lambda_1=0.3 \sqrt{2}$. The white dashed line represents the second-order boundary $\gamma = 1/\lambda_+$ which terminates at the TP as represented by the white star. }
\label{Fig3}
\end{figure}

Although $H_0$ is enough to determine the ground state mean-field properties, further terms ($H_1$ and $H_2$) are needed to study the excitation spectrum. As shown in Appendix~\ref{AppA}, for any values of the order parameters $\alpha$, $\beta_1$, and $\beta_2$ that minimize $H_0$, the Hamiltonian $H_1$ vanishes, and the excitation spectrum is determined by $H_2$. The general form of $H_2$ is given by:
\begin{equation}
    H_2 = \sum_{j=1}^6 \sum_{k=1}^6 \mathcal{C}_{jk} v_j v_k\,,
    \label{H2}
\end{equation}
where $v_j$ is the $j$-th component of the operator vector $\vec{v}=(c^{\dagger},d_1^{\dagger},d_2^{\dagger},c,d_1,d_2)$, and the matrix components $\mathcal{C}_{jk}$ are given explicitly in Appendix~\ref{AppA}. Since the Hamiltonian in Eq.~(\ref{H2}) is bilinear in the annihilation and creation operators, it can be diagonalized using a Bogoliubov transformation \cite{Bogoljubov1958-te,valatin1958comments} (see Appendix~\ref{AppC}) into the form
\begin{equation}
    H_2 = \sum_{j=1}^3 \varepsilon_j a_j^{\dagger} a_j \, ,
    \label{H2bogo}
\end{equation}
where we have omitted a constant shift. The annihilation and creation operators $a_j$ and $a_j^{\dagger}$ are a linear combination of all the operators contained in the components of $\vec{v}$. If we consider that $\varepsilon_1 < \varepsilon_2 < \varepsilon_3$ for a given set of all system parameters, then we can identify $\varepsilon_1 = \Delta$ as the energy gap between the ground state and the first excited state. In a second-order phase transition, including the TP, we expect $\Delta$ to vanish exactly at the phase transition, this is illustrated in Fig.~\ref{Fig3}.

Note that the second order equation in Eq.~(\ref{TPcond1}) agrees with the numerical behavior as signaled by the white dashed line. When the first order line is crossed, a discontinuous jump in the energy gap is observed. Note that here we chose to illustrate the energy gap variation entering the SRA phase. However, as evidenced by the symmetry of the phase diagrams in Fig.~\ref{Fig2} an identical behavior is expected for the SRB phase. We note that although Eq.~(\ref{H2bogo}) contains multiple possible excitations, we only consider the first excited state here as higher excited states might not be described correctly using the generalized Holstein-Primakoff map \cite{Hirsch_2013}. While not explored here, the description of higher energy states could be done for a finite number of atoms $N$ through exact diagonalization following the method described in the next section.

To differentiate between the different types of phase boundaries we can explore their corresponding critical exponents. For instance, let us consider a point $p=(\delta, \lambda_1, \lambda_2, \gamma)$ located very close to the critical point $p_c = (\delta_c,\lambda_{1c},\lambda_{2c},\gamma_c)$, formally, we consider $p$ to be located in a line perpendicular to the phase boundary at point $p_c$. We expect that the order parameter $\alpha$ scales as $\alpha \propto d^{\mu}$, where $d=\sqrt{(\delta - \delta_c)^2+(\lambda_1 - \lambda_{1c})^2+(\lambda_2 - \lambda_{2c})^2+(\gamma - \gamma_c)^2}$ is the distance from the critical point when we approach it from the superradiant phase. Similarly, we could define the scaling behavior of the excitation gap $\Delta \propto d^{\nu_{\pm}}$, where $\nu_-$ considers the point $p$ to be located in the normal phase and $\nu_+$ is the scaling exponent when the boundary is approached from the superradiant phase. 

\begin{figure}[t!]
\includegraphics[width=0.48\textwidth]{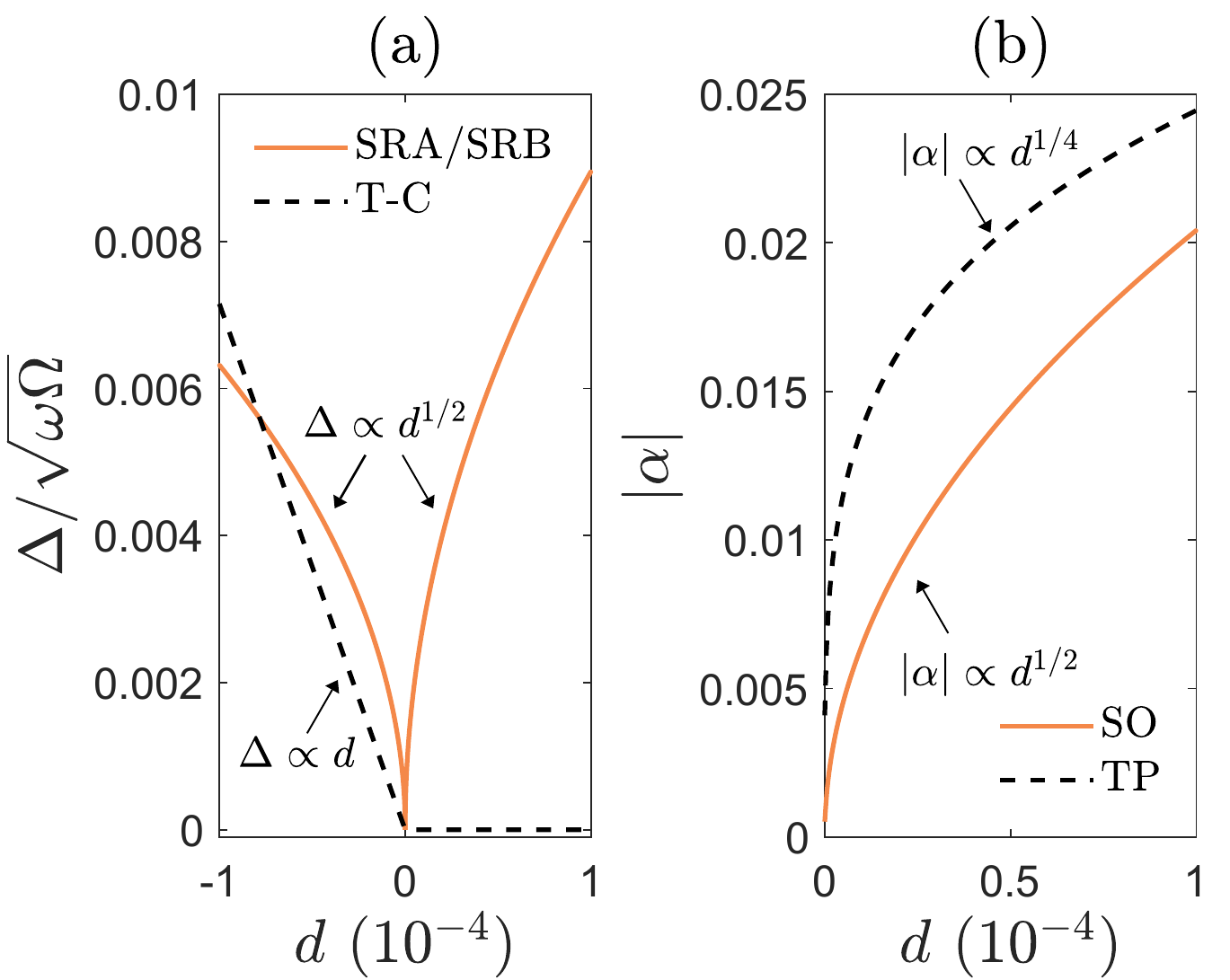}
\caption{Scaling near different types of transitions. (a) The energy gap near the critical point. The dashed line corresponds to the Tavis-Cummings model where the counter-rotating terms are not present. The solid line depicts the behavior when both co- and counter-rotating terms are present. (b) The order parameter $\vert \alpha \vert$ near the critical point. The dashed (solid) line represents the behavior across the TP (a second-order critical point). We consider values of $\vert d \vert \leq 1 \times 10^{-4}$, and $d$ is defined to be negative (positive) if the transition is approached from the normal (superradiant) phase. The scaling exponent for each phase transition is signaled with an arrow.}
\label{Fig4}
\end{figure}

In Fig.~\ref{Fig4}, the scaling behavior of $\Delta$ and $\vert \alpha \vert$ is presented. The first thing it must be noted is that the SRA and SRB phases have identical scaling exponents, this means that in an experiment, where the available quantity to measure is $\vert \alpha \vert^2$, these phases are indistinguishable. Regardless of crossing a second-order boundary or a TP, the excitation gap vanishes with $\nu_{\pm}=1/2$ for these two phases. For the Tavis-Cummings line, on the other hand, as we reach the critical point (second order or TP) from the NP, the energy gap vanishes with $\nu_- = 1$. However, the energy gap remains equal to zero inside the superradiant phase as shown in Fig.~\ref{Fig4}(a). This Goldstone mode \cite{goldstone} is characteristic of phases where a continuous $U(1)$ symmetry is spontaneously broken, leading to these gapless excitations.

In the same way that $\nu_{\pm}$ can be used to differentiate between the Tavis-Cummings line and the SRA/SRB phase transitions, the exponent $\mu$ allows differentiating between a second-order line and a TP as shown in Fig.~\ref{Fig4}(b). After crossing to any superradiant phase the order parameter scales with $\mu=1/2$ for a second order phase transition while the exponent is $\mu=1/4$ for a TP, indicating that TP belongs to a different universality class in comparison to other critical points on the second-order line. 

\section{Finite $N$}

Now that all the mean-field features in the thermodynamic limit of the model have been discussed, it is important to study if the precursors of the phase transitions are still present for a finite number of atoms $N$. In order to perform exact diagonalization calculations we need to consider a cutoff photon number $N_{ph}$, this means that the total size of the Hilbert space is $3^{N}\times N_{ph}$. Clearly, only a few atoms can be considered if the full Hilbert space is used. However, as we will show below, by exploiting symmetry constraints, we are able to consider a much larger system with $N \sim 10^2$. 

To this end, we can borrow some ideas from the treatment used in the conventional Dicke model, see for example Ref.~\cite{Emary2003}. The conventional Dicke Hamiltonian is given by $H=\omega a^{\dagger}a + \Omega S_z + g/\sqrt{N}(a+a^{\dagger})S_x$, where $S_j$ are collective spin operators. It is clear that $[H,S^2]=0$, which means that $S^2$ is conserved and that states with different eigenvalue $S$ are not mixed by the Hamiltonian. Since in the mean-field ground state of the NP, all spins point downwards (i.e., all the atoms are in the ground state), it is of interest to consider the totally symmetric manifold with $S=N/2$ to which the NP ground state belongs, and to represent the Hamiltonian using only the set of states $\{ \vert N/2, -N/2 \rangle, \vert N, -N/2+1 \rangle ,...,\vert N/2, N/2 \rangle \}$. These states $|S=N/2,m \rangle$ are often referred to as Dicke States and using them reduces the atomic Hilbert space size from $2^N$ to $N+1$. Our task is then to find the corresponding Dicke states for the TDM.

\subsection{Gell-Mann matrices}
The TDM Hamiltonian in Eq.~(\ref{TDMH}) clearly does not commute with $S^2$ as it is nonlinear in the spin operators $S_j$. However, if instead of considering an $SU(2)$ representation through the conventional spin operators, we choose an $SU(3)$ representation spanned by the Gell-Mann matrices $\Lambda_j$ (see Appendix~\ref{AppD} for a list of the Gell-Mann matrices' properties), the TDM Hamiltonian can be recast in the following form which is linear in terms of the Gell-Mann matrices:
\begin{eqnarray}
    H &=& \omega a^{\dagger}a + \frac{\Omega}{2} \left( \frac{3-\delta}{\sqrt{3}} \Lambda_8+ (1-\delta)\Lambda_3\right) \nonumber \\
    &&+\frac{g_1}{2\sqrt{N}} a(\Lambda_1 + i \Lambda_2+\gamma \Lambda_6 + i \gamma \Lambda_7) \nonumber \\
    &&+\frac{g_1}{2\sqrt{N}} a^{\dagger}(\Lambda_1 - i \Lambda_2+\gamma \Lambda_6 - i \gamma \Lambda_7) \nonumber \\
    &&+\frac{g_2}{2\sqrt{N}} a^{\dagger}(\Lambda_1 + i \Lambda_2+\gamma \Lambda_6 + i \gamma \Lambda_7) \nonumber \\
    &&+\frac{g_2}{2\sqrt{N}} a(\Lambda_1 - i \Lambda_2+\gamma \Lambda_6 - i \gamma \Lambda_7) \, . 
    \label{TDMHGM}
\end{eqnarray}
Just as in Eq.~(\ref{TDMH}), we have summed over all atoms and written the Hamiltonian in terms of collective operators, namely, $\Lambda_j = \sum_{k=1}^N \Lambda_j^{(k)}$. 
One can show that Hamiltonian (\ref{TDMHGM}) commutes with the two Casimir operators of $SU(3)$:
\begin{equation}
C_1 = \sum_j \Lambda_j \Lambda_j, \quad C_2 = \sum_{j,k,l} d_{jkl} \Lambda_j \Lambda_k \Lambda_l \,,
\label{casimir}
\end{equation}
where $d_{jkl}=\frac{1}{4}\textnormal{tr}(\{\Lambda_j,\Lambda_k\}\Lambda_l)$ are totally symmetric coefficients. A conventional approach is to use the Cartan-Weyl notation instead of the Gell-Mann matrices, so we define:
\begin{eqnarray}
&T_{\pm} = \frac{1}{2}(\Lambda_1 \pm i \Lambda_2), \quad T_z = \frac{1}{2}\Lambda_3, \quad Y = \frac{1}{\sqrt{3}} \Lambda_8 \,, \nonumber \\
&U_{\pm} = \frac{1}{2}(\Lambda_6 \pm i \Lambda_7), \quad V_{\pm} = \frac{1}{2}(\Lambda_4 \pm i \Lambda_5) \, .
\label{cartanweyl}
\end{eqnarray}
In this notation, it is clear that there are three sets of ladder operators driving the transitions between the three different states, while $T_z$ and $Y$ are both diagonal operators and are associated with isospin and hypercharge in the context of particle physics \cite{gellmann}. 

In terms of the operators defined in Eqs.~(\ref{cartanweyl}) and up to a constant shift, the TDM Hamiltonian becomes:
\begin{eqnarray}
    H &=& \omega a^{\dagger}a + \Omega \left( \frac{3-\delta}{2}Y + (1-\delta)T_z \right) \nonumber \\
    &&+\frac{g_1}{\sqrt{N}}(a (T_++\gamma U_+)+a^{\dagger}(T_-+\gamma U_-)) \nonumber \\
    &&+\frac{g_2}{\sqrt{N}}(a^{\dagger} (T_++\gamma  U_+)+a(T_-+\gamma U_-))\,. 
    \label{TDMHCW}
\end{eqnarray}

\subsection{$SU$(3) Dicke states}
Similar to how different representations of $SU(2)$ are labeled by the different eigenvalues of $S^2$, different representations of $SU(3)$ will be classified depending on the eigenvalues of the two Casimir operators $C_1$ and $C_2$, which we denote as $c_1$ and $c_2$, respectively. A common notation change is to consider the integers $p$ and $q$ instead of $c_1$ and $c_2$ as the labels for the representations. The relation between these two notations is given by \cite{pais}:
\begin{eqnarray}
    c_1 &=& (p^2+q^2+3p+3q+pq)/3 \,, \nonumber\\
    c_2 &=& (p-q)(3+p+2q)(3+q+2p)/18 \, .
\end{eqnarray}
In the particle physics context, $p$ and $q$ correspond to the number of quarks and antiquarks, respectively \cite{greiner2012quantum}.
Since $Y$ and $T_z$ commute with each other, they can define a set of commutable operators with $C_1$ and $C_2$. Moreover, if we define $T^2 = T_x^2+T_y^2+T_z^2$, with $T_\pm = T_x \pm i T_y$, the set $\{ T^2,T_z,  Y, C_1, C_2 \}$ defines a complete set of commutable operators \cite{baird1963}. Consequently, each state in a given representation is labeled by the eigenvaleus of these operators, namely,  $\vert t,t_z,y,p,q \rangle$. 

Since the TDM Hamiltonian commutes with $C_1$ and $C_2$, it does not mix states with different values of $p$ and $q$. Similar to the conventional Dicke states, we focus on the totally symmetric representation given by $q=0$, $p=N$ \cite{greiner2012quantum}. As shown in Ref.~\cite{macfarlane1963weyl}, in the totally symmetric representation, $y$ and $t$ are related by $t=y/2 +p/3$. Then, we can omit the labels $p$, $q$, and $y$, and the states of interest are simply labeled by $\vert t, t_z \rangle$, with $t=0,\,1/2,\,1,\,3/2,...,\,N/2$ and $t_z = -t,\,-t+1,...,\,t-1,\,t$.  These states represent the generalized Dicke states~\cite{Rosso2022,hartmann} for $SU(3)$. Different $SU(3)$ Dicke states are connected by the ladder operators $T_{\pm}$ and $U_\pm$.

Once these $SU(3)$ Dicke states are chosen as a basis, the only thing missing is to find the matrix elements of relevant operators under this basis in order to construct the matrix representation for Hamiltonian~(\ref{TDMHCW}). The explicit expressions for these operators can be found in Appendix \ref{AppE}. As $T_z$, $T_+$, and $T_-$ define an $SU(2)$ subalgebra, their matrix elements are very easy to determine. By contrast, $U_{\pm}$ produce interesting matrix elements as they change the value of both $t$ and $t_z$, simultaneously. 

The dimension of the totally symmetric subspace spanned by the $SU(3)$ Dicke states is $(N+1)(N+2)/2$, which means that we have decreased the atomic Hilbert space size from being exponential in $N$ to quadratic in $N$. Furthermore, the parity operator $\Pi = \textnormal{exp}(i \pi \left(a^{\dagger}a + T_z + 3Y/2 \right))$ commutes with the Hamiltonian~(\ref{TDMHCW}). Hence this symmetric subspace can be further divided into two: one with even and the other with odd parity. Consequently, the size of the Hilbert space needed for exact diagonalization is reduced by another factor of 2. 

\begin{figure}[t!]
\includegraphics[width=0.48\textwidth]{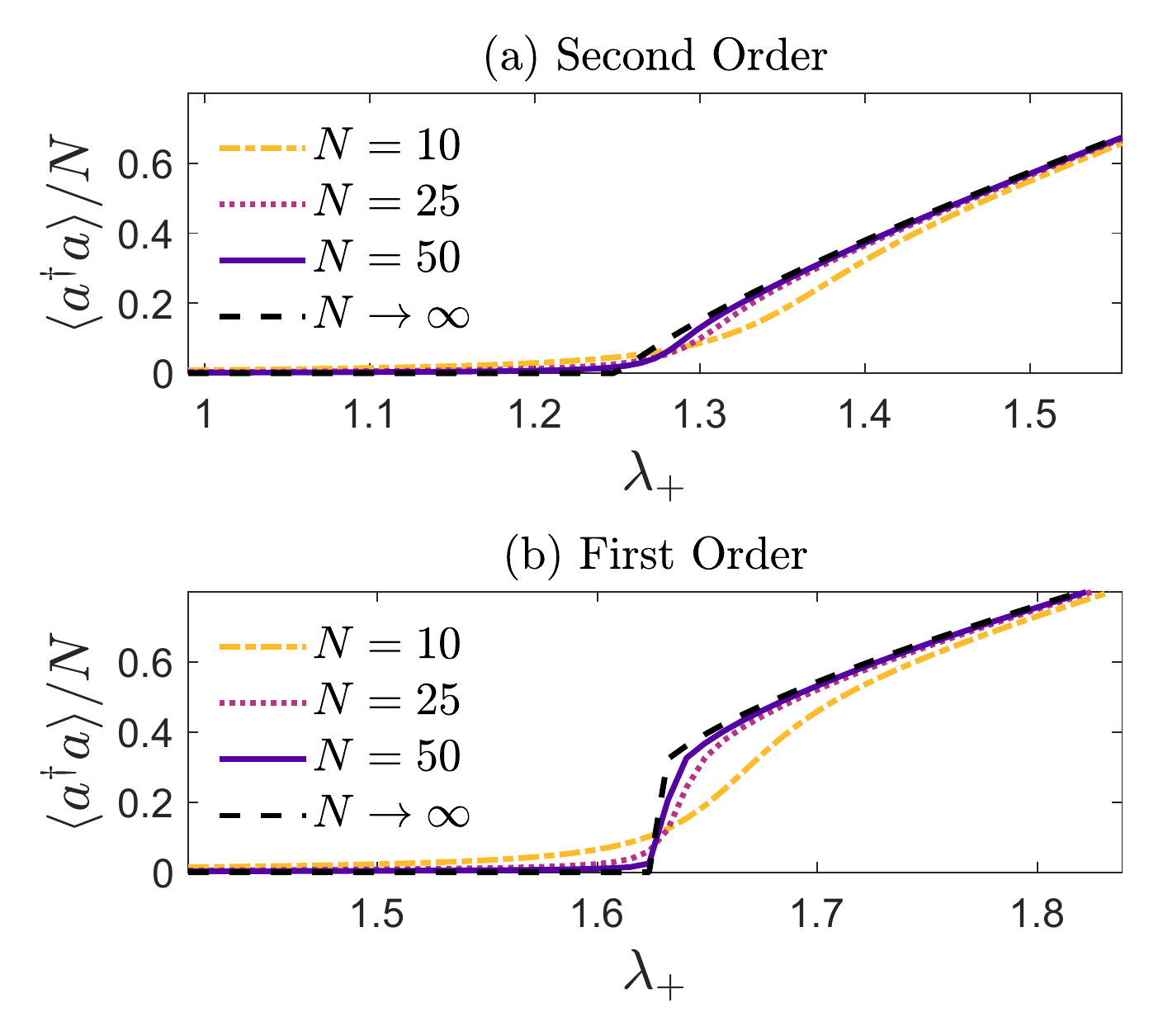}
\caption{Photon population as a function of $\lambda_+$ for different atom number $N$ near a phase transition into the SRA phase. In (a) the transition is of second order with $\gamma=0.8$, while in (b) the transition is of first order with $\gamma=0.6$. In both cases, we consider $\lambda_1=\lambda_2=\lambda_+/2$ and $\delta=0$. The dashed line shows the mean field behavior in the thermodynamic limit. We consider a photon cutoff of $N_{ph}=100$.}
\label{Fig5}
\end{figure}

In Fig.~\ref{Fig5}, the behavior of the photon population $\langle a^{\dagger} a \rangle$ for a finite number of atoms $N$ is compared with the results from the thermodynamic limit $N \rightarrow \infty$. Note that we cannot compare $\langle a \rangle$ since the spontaneous symmetry breaking only occurs in the thermodynamic limit, namely, for finite $N$ it is always the case that $\langle a \rangle =0$.

We note from the figure that as $N$ increases the behavior of the photon population converges rapidly to the expected behavior in the thermodynamic limit. Moreover, both the smooth behavior of the second order phase transition and the sharp discontinuous behavior of the first order line can already be captured with $N=50$. Then, in an experimental realization where thousands of atoms could be trapped, we expect that the phase transition could be easily characterized by the behavior of $\langle a^{\dagger}a \rangle/N$.

The convergence of the results as $N$ increases signals that the description using the generalized Holstein-Primakoff map, and the one using the Gell-Mann matrices are equivalent to each other in the appropriate limit $N \rightarrow \infty$.

\section{Open system steady states}

A potential experimental realization of the Hamiltonian~(\ref{TDMHCW}) can be done using the hyperfine states of an atom through cavity-assisted Raman transitions, this was proposed for realizing spin-1 light-interacting Hamiltonians in Ref.~\cite{Masson2017}, realized experimentally in Ref.~\cite{Zhiqiang:17}, and could be extended to higher spin systems as proposed in Ref.~\cite{Youjiang2021}, see Appendix~\ref{AppF} for more details. In practical situations involving cavities, dissipative processes are unavoidable. This raises the question of whether all the types of critical boundaries that we find in equilibrium would survive once incoherent losses are taken into account. In particular, how would the TP manifest in an open system?

We focus only on the leaking of photons out of the cavity with rate $\kappa$ as in the considered setup we can omit atomic spontaneous decay (see Appendix~\ref{AppF}). In the absence of counter-rotating terms (Tavis-Cummings line), even an infinitesimal value of $\kappa$ would suppress the dissipative phase transition~\cite{Soriente}. Hence we focus exclusively on the dissipative phase transition into the SRA. For simplicity, we consider $g_1=g_2=g$ and $\delta=0$. This means that $\lambda_1=\lambda_2$ and the light matter interaction is reduced to a single parameter, to keep a consistent notation we choose that parameter to be $\lambda_+ = 2\lambda_1 = 2\lambda_2$. 

\subsection{Master equation}

Since the complete expressions for $C_1$ and $C_2$ in Eq.~(\ref{casimir}) will be used as constraints, in this case, it is simpler to consider the Hamiltonian in terms of the Gell-Mann Matrices as in Eq.~(\ref{TDMHGM}).
The open system dynamics are described by the Heisenberg picture's Lindblad equation:
\begin{equation}
\frac{d}{dt} \mathcal{A} = i [H, \mathcal{A}] + \kappa (2 a^{\dagger}\mathcal{A}a -\{a^{\dagger}a,\mathcal{A} \})\,,
\label{AdjLindblad}
\end{equation}
where $\mathcal{A}$ represents any operator of interest. We can obtain a system of coupled differential equations by computing Eq.~(\ref{AdjLindblad}) for all $\Lambda_i$'s and $a$:
\begin{eqnarray}
        \frac{d}{dt} \langle a \rangle &=& -i (\omega - i \kappa) \langle a \rangle  - i g (\langle \Lambda_1 \rangle + \gamma \langle \Lambda_6 \rangle) \nonumber \\
        \frac{d}{dt} \langle \Lambda_1 \rangle &=& - \Omega \langle \Lambda_2 \rangle + g \gamma (\langle a \rangle+ \langle a^{\dagger} \rangle )\langle \Lambda_5 \rangle \nonumber \\
    \frac{d}{dt} \langle \Lambda_2 \rangle &=&  \Omega \langle \Lambda_1 \rangle - 2 g (\langle a \rangle +\langle a^{\dagger} \rangle) \langle \Lambda_3 \rangle \nonumber \\
   & &- g \gamma (\langle a \rangle +\langle a^{\dagger} \rangle) \langle \Lambda_4 \rangle \nonumber \\
    \frac{d}{dt} \langle \Lambda_3 \rangle &=&  2g (\langle a \rangle+ \langle a^{\dagger} \rangle) \langle \Lambda_2 \rangle - g\gamma(\langle a \rangle+ \langle a^{\dagger}\rangle) \langle \Lambda_7 \rangle \nonumber\\
    \frac{d}{dt} \langle \Lambda_4 \rangle &=&  -2 \Omega \langle \Lambda_5 \rangle - g(\langle a \rangle +\langle a^{\dagger}\rangle) \langle \Lambda_7 \rangle \nonumber \\
    &&+g \gamma(\langle a \rangle + \langle a^{\dagger} \rangle) \langle \Lambda_2 \rangle\nonumber \\
    \frac{d}{dt} \langle \Lambda_5 \rangle &=&  2 \Omega \langle \Lambda_4 \rangle + g (\langle a \rangle +\langle a^{\dagger} \rangle) \langle \Lambda_6 \rangle \nonumber \\
    &&-g\gamma(\langle a \rangle +\langle a^{\dagger} \rangle)\langle \Lambda_1 \rangle \nonumber \\
    \frac{d}{dt}\langle \Lambda_6 \rangle &=&  -\Omega \langle \Lambda_7 \rangle - g(\langle a \rangle +\langle a^{\dagger} \rangle ) \langle \Lambda_5 \rangle \nonumber \\
    \frac{d}{dt} \langle \Lambda_7 \rangle &=&   \Omega \langle \Lambda_6 \rangle + g (\langle a \rangle +\langle a^{\dagger} \rangle) \langle \Lambda_4 \rangle \nonumber \\
    &&+ g \gamma (\langle a \rangle +\langle a^{\dagger}\rangle ) \langle \Lambda_3 \rangle    \nonumber \\
    && - \sqrt{3} g \gamma (\langle a \rangle +\langle a^{\dagger} \rangle )\langle \Lambda_8 \rangle \nonumber \\
    \frac{d}{dt} \langle \Lambda_8 \rangle &=&  \sqrt{3}g \gamma (\langle a \rangle + \langle a^{\dagger} \rangle ) \langle \Lambda_7 \rangle\,,
\label{systemofODES}
\end{eqnarray}
where we have taken the expectation value on both sides of each equation. Note that we have taken the mean field approximation where expectation values of the form $\langle a \Lambda_i \rangle $ are approximated by $\langle a \rangle \langle \Lambda_i \rangle$. This approximation has proven to be very effective in open Dicke-like systems when  working in the thermodynamic limit $N \rightarrow \infty$ \cite{Bhaasen}. Consequently, for all the following results we will always consider the system in the thermodynamic limit. We have also rescaled the expectation values as $\langle \Lambda_i \rangle /N \rightarrow \langle \Lambda_i \rangle $ and $\langle a \rangle /\sqrt{N} \rightarrow \langle a \rangle $. 

\begin{figure*}[t!]
\includegraphics[width=0.95\textwidth]{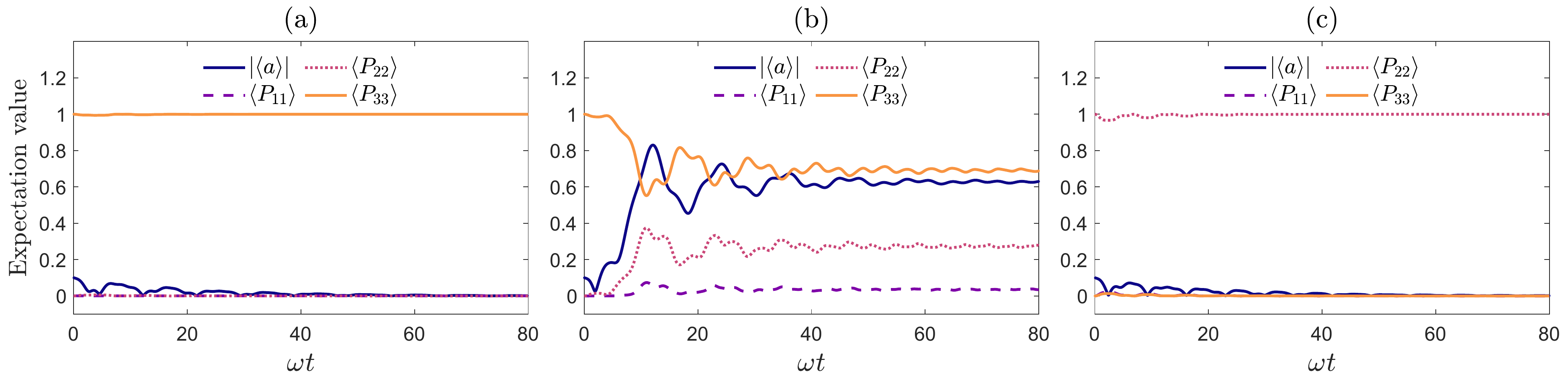}
\caption{Steady state stability. In (a) the system parameters are fixed to $\gamma=0.8$ and $\lambda_+ = 0.6\sqrt{2}$, while in (b) and (c) the parameters are set to $\gamma=0.8$ and $\lambda_+ = \sqrt{2}$. In all panels the state is initialized in a normal phase but with a slightly perturbed initial photon population $\langle a \rangle = 0.1 + 0.01i$. In (a) and (b) the initial state is very close to the NP3 phase while in (c) the initial state is very close to the NP2 phase. In all panels $\kappa/\omega=0.1$.}
\label{Fig6}
\end{figure*}

Since we will focus only on the steady-state properties of the system, we set all equations in Eq.~(\ref{systemofODES}) equal to zero. Two important results follow from the steady-state equations. First, the steady state expectation value of all the antisymmetric $\Lambda_i$ operators vanishes, namely, $\langle \Lambda_2 \rangle = \langle \Lambda_5 \rangle = \langle \Lambda_7 \rangle =0$; and second, although there are still seven real variables to determine (note that $\langle a \rangle$ counts as two variables as it is generally complex), only five independent equations remain. After algebraic manipulation of Eqs.~(\ref{systemofODES}), it can be shown that the two additional constraints are given by:
\begin{eqnarray}
    A = \sum_j \langle \Lambda_j \rangle \langle \Lambda_j \rangle, \quad B = \sum_{j,k,l} d_{jkl} \langle  \Lambda_j \rangle \langle  \Lambda_k \rangle \langle \Lambda_l \rangle \,,
    \label{extracons}
\end{eqnarray}
where $A$ and $B$ are time independent, i.e., $dA/dt=0=dB/dt$. Hence $A$ and $B$ are two constants determined by the initial conditions. It is clear that these two additional constraints are a manifestation of the Casimir invariants in Eq.~(\ref{casimir}). This is a similar situation to what happens in the conventional open Dicke model where the extra constraint needed arises from the conservation of the total spin length ($SU(2)$ Casimir invariant). In the thermodynamic limit, and in the totally symmetric representation where $p=N$, the eigenvalues $c_1$ and $c_2$ are given by:
\begin{equation}
    \frac{c_1}{N^2} \approx \frac{1}{3}, \quad \frac{c_2}{N^3} \approx \frac{1}{9}\,.
\end{equation}
Since rescaling $\langle \Lambda_j \rangle \rightarrow D \langle \Lambda_j \rangle$, with $D$ a time-independent constant, does not change Eq.~(\ref{extracons}), we can define $A= D^2 c_1 / N^2$ and $B = D^3 c_2/N^3$. Here we choose $D=2$ such that $A=4/3$ and $B=8/9$. Now that we have a complete set of algebraic equations we can solve for all possible steady states which can be broadly divided into four categories (three normal phases and one superradiant phase) as shown in Table \ref{classes}.

\begin{table}[t]
\caption{Different steady state phases.}
\begin{tabularx}{0.45\textwidth} { 
  | >{\centering\arraybackslash}X 
  | >{\centering\arraybackslash}X|}
 \hline
Phase & Expectation values\\
 \hline
 Normal phase 1 (NP1) & $\langle a \rangle =0$, $\langle P_{11} \rangle = 1$ $\langle \Lambda_1 \rangle = \langle \Lambda_4 \rangle = \langle \Lambda_6 \rangle =0$\\
 \hline
 Normal phase 2 (NP2)& $\langle a \rangle =0$, $\langle P_{22}\rangle = 1$  $\langle \Lambda_1 \rangle = \langle \Lambda_4 \rangle = \langle \Lambda_6 \rangle =0$\\
\hline
 Normal phase 3 (NP3) & $\langle a \rangle =0$, $\langle P_{33}\rangle = 1$ $\langle \Lambda_1 \rangle = \langle \Lambda_4 \rangle = \langle \Lambda_6 \rangle =0$\\
\hline
Superradiant phase (SR) & $\langle a \rangle \neq 0$, $\langle \Lambda_1 \rangle$,$\langle \Lambda_4 \rangle$,$\langle \Lambda_6 \rangle \neq 0$ \\
\hline
\end{tabularx} 
\label{classes}
\end{table}

Although we find four categories of steady states, it does not mean that all of them are stable attractors. In order to study the stability of each steady state, we can simulate the dynamics of the system of differential equations in Eq.~(\ref{systemofODES}) starting from slightly perturbed states and check if the dynamics lead the system back to this same steady state. This is illustrated in Fig.~\ref{Fig6} where we initialize the system in a slightly perturbed state with respect to different normal phases. For the parameters in Fig.~\ref{Fig6}(a), the NP3 phase is stable and the system rapidly goes back to this state after it is slightly perturbed. By contrast, in (b) the perturbation causes the system to evolve away from the unstable NP3 phase into a stable SR phase. 

\subsection{Dissipative phase diagram}

\begin{figure}[t!]
\includegraphics[width=0.48\textwidth]{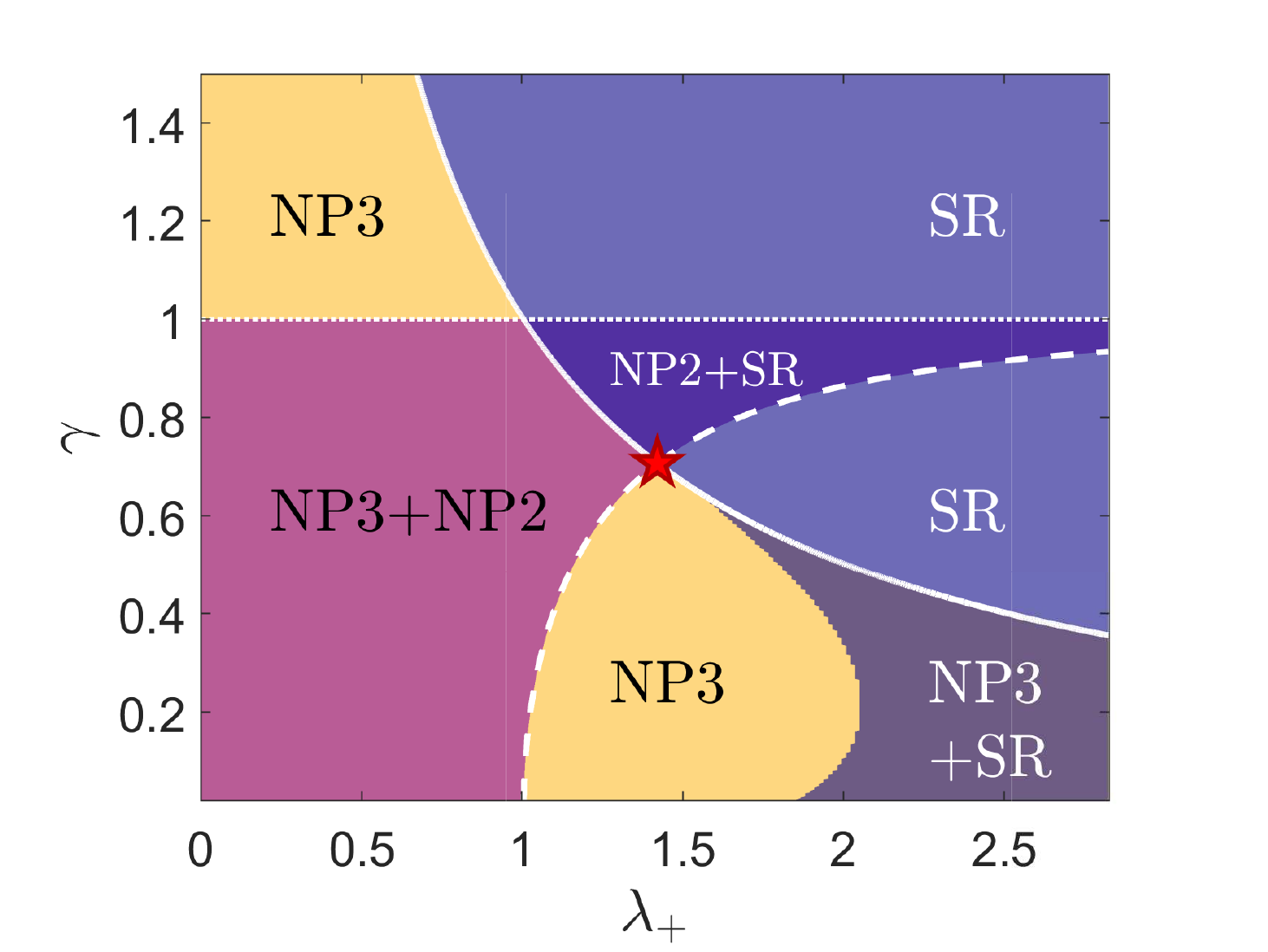}
\caption{Phase diagram of the stable steady states in the $\lambda_+$-$\gamma$ plane for $\kappa/\omega = 0.1$. The white solid line represents the stability boundary of the NP3 phase, while white dashed and dotted lines bound the stability region of the NP2 phase. The TP, signaled by a red star, is located at the intersection of two of the stability boundaries.}
\label{Fig7}
\end{figure}

In Fig.~\ref{Fig7}, a phase diagram with all the stable steady states is presented for $\kappa/\omega=0.1$. The first key thing to notice is that while the NP1 phase is always unstable, both NP2 and NP3 have regions where they are stable. Specifically, we note that for $\gamma > 1$, the NP2 phase is always unstable regardless of the value of $\lambda_+$. This behavior can be intuitively understood using the schematics in Fig.~\ref{Fig1}. If $\gamma > 1$, the coupling for the lower atomic transition between states $\vert 3 \rangle$ and $\vert 2 \rangle$ is stronger than the coupling for ther upper transition between state $\vert 2 \rangle$ and $\vert 1 \rangle$. Since state $\vert 1 \rangle$ has the highest energy of all, it follows that in the normal phase the system behaves like an effective two-level system and, as in the conventional Dicke model, in which the only stable normal phase is that where all spins populate the lowest energy state, in this case, state $\vert 3\rangle$. 

For $\gamma < 1$, on the other side, all the richness of having three-level atoms can be exploited and we see a series of different stability regions, of particular interest are the regions of bistability where two different phases are stable and the final fate of the system would depend entirely on the initial conditions. These bistable regions can contain two normal phases or one normal phase and one superradiant phase. The dependence on initial conditions in a bistable region is illustrated in Fig.~\ref{Fig6} (b) and (c) where, for the same set of parameters, different initial states lead the system to the superradiant phase in (b) and to the NP2 phase in (c).

The stability of each steady state can be examined using the standard linear stability analysis. Using the generalized Holstein-Primakoff mapping and by considering the small fluctuations above the steady state, the stability boundaries of the NP2 and NP3 phases can be found analytically (see Appendix~\ref{AppG} for more details) as:
\begin{eqnarray}
    &\lambda_+ = \frac{\sqrt{1+\kappa^2/\omega^2}}{\gamma} \rightarrow {\rm NP}3 \,,\nonumber \\
    &\lambda_+ = \frac{\sqrt{1+\kappa^2/\omega^2}}{\sqrt{1-\gamma^2}} \rightarrow {\rm NP}2 \,.
    \label{stabilitybounds}
\end{eqnarray}
These two boundaries are represented by the solid and dashed lines, respectively, in Fig.~\ref{Fig7}. These two lines intersect at the point \begin{equation} \lambda_{\rm +TP}=\sqrt{2+2\kappa^2/\omega^2},\,\;\;\;\gamma_{\rm TP}=1/\sqrt{2}\,, \label{tpopen}
\end{equation} 
represented by the red star in Fig.~\ref{Fig7}. This is the only point where all three bistability regions converge. In fact, it is the point where all the steady state phases for $\gamma<1$ converge. As we shall argue below, this special point is the TP in the open system.

\subsection{Tricritical behavior}
As bistability is often associated with first-order phase transition, the emergence of bistable regions studied above is intimately connected to the TP in the closed system. In the limit of $\kappa \rightarrow 0$, the values of $\lambda_+$ and $\gamma$ coincide exactly with the values of the TP in the closed system ({\color{red}white star in Fig.~\ref{Fig3}}). The effect of cavity decay is to increase the critical coupling strength to a larger value. A similar behavior is also seen in the conventional open Dicke model.
On the other hand, since the leaking of photons affects both the upper and the lower atomic transitions in the same manner, it is reasonable to argue that the critical value of $\gamma$, which characterizes the ratio of the coupling strengths for the two transitions, for the open system TP should remain unchanged with respect to the closed system value. These qualitative arguments strongly suggest that the red star in Fig.~\ref{Fig7} is indeed the TP in this open system. 

\begin{figure}[t!]
\includegraphics[width=0.48\textwidth]{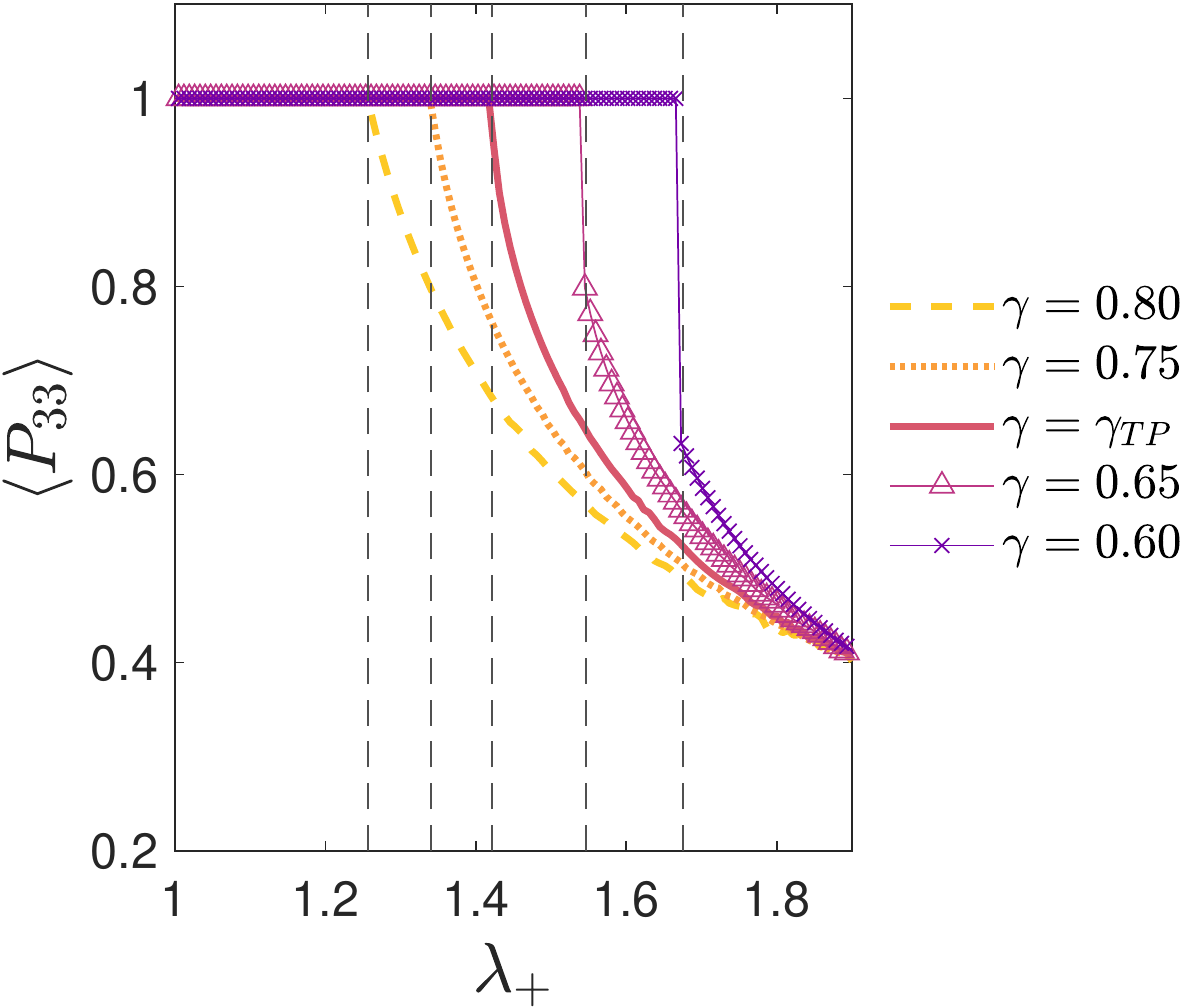}
\caption{Steady state value of the population on state $\vert 3 \rangle$ as a function of $\lambda_+$ for different values of $\gamma$. Each value of $\langle P_{33} \rangle$ is found by numerically integrating the set of differential equations in Eq.~(\ref{systemofODES}) for a very long time $t \omega = 10000$. For all data points the initial state is slightly perturbed from the NP3 phase with $\langle a \rangle = 0.1 + 0.01i$. The vertical dashed gray lines represent the value of $\lambda_+$ where we expect the NP3 phase to become unstable for each  $\gamma$ according to Eq.~(\ref{stabilitybounds}). We set $\kappa/\omega = 0.1$ for all points.}
\label{Fig8}
\end{figure}

In order to quantitatively characterize the tricritical behavior in the open system and compare it to that of the closed system where we only have the NP3 and SR phases, we consider horizontal cuts (fixed $\gamma$) of the phase diagram in Fig.~\ref{Fig7} and vary $\lambda_+$. For each value of $\lambda_+$ we start the dynamics in a state slightly perturbed from the NP3 phase and then let the system evolve until it reaches the steady state. The steady-state values of the population in state $\vert 3 \rangle$, namely, $\langle P_{33}\rangle$ are shown in Fig.~\ref{Fig8}.

We note that as we sweep $\lambda_+$ for $\gamma \geq \gamma_{\rm TP}$, the change of the NP3 phase going from stable to unstable is signaled by a continuous change in $\langle P_{33} \rangle$ from one into a smaller value, a behavior expected from a second-order transition. For $\gamma \leq \gamma_{\rm TP}$, by contrast, the change in $\langle P_{33} \rangle$ is discontinuous as we would expect in a first-order transition. This change from continuous to discontinuous behavior confirms the nature of the tricritical point. Note that since all points in Fig.~\ref{Fig8} are obtained from an initial state very close to the NP3 phase, then it is clear that the critical points should follow Eq.~(\ref{stabilitybounds}), obtained using the generalized Holstein-Primakoff mapping, as illustrated by the gray vertical dashed lines. This again shows the excellent consistency between the generalized Holstein-Primakoff mapping and the use of the Gell-Mann matrices. In a similar fashion, by choosing different initial states we could take a look at all the other available steady states.

As far as we know, ours is the first study on how a TP in a closed system manifests itself when dissipation is included, although Overbeck {\it et al.} previously investigated the TP in a dissipative Ising model~\cite{triIsing1} which does not exist in the absence of dissipation. Nevertheless, a more general theory of multicriticality in open systems is still lacking and should be an interesting direction for future studies.

\section{Conclusion}

We have presented a thorough study of the tricritical Dicke model in both the closed and the open setups. In equilibrium, the tunability of the system allows studying not only the change of the phase transition order from second to first but also the spontaneous symmetry breaking of both discrete and continuous symmetries. The different phase transitions were classified according to their scaling exponents. Moreover, signals of these transitions were shown to be observable far from the thermodynamic limit (with less than a hundred atoms).

In the presence of cavity losses, the system develops a series of regions of bistability, the emergence of which is closely related to the tricritical point, and all bistable regions converge at the tricritical point. Additionally, the NP2 phase becomes stable in a large region of the parameter space for $\gamma \leq \gamma_{\rm TP}$. The richness of the non-equilibrium phase diagram allows for the potential preparation of desired steady states through proper choices of initial states and/or parameter quenching.

Both in the closed and the open setups there is an excellent agreement between using the generalized Holstein-Primakoff mapping and the description using the Gell-Mann matrices in the appropriate limit $N \rightarrow \infty$. Nonetheless, more than just being equivalent, the two approaches are complementary as different levels of information about the system can be accessed through each one of them.

Although this system could be realized using Raman transitions as mentioned above, an interesting future research direction could be to use other platforms where an equivalent system might be realized to explore these interesting steady states. There are already various platforms where a Dicke-like Hamiltonian can be simulated \cite{Baumann2010-zm,li2018,Mezzacapo2014-xl}, and modifications in those setups might lead to Hamiltonians of the form of Eq.~(\ref{TDMH}). For example, in a spin-orbit coupled BEC with spin-1 atoms, tricritical points have been reported \cite{campbell2016magnetic}. In that case, the motional degrees of freedom of the atoms are the analog of the light mode in our setup \cite{dickesoc12345}. Then, if a loss process equivalent to the photons leaking from the cavity can be engineered in such a platform, various interesting magnetic steady states could be explored.

\begin{acknowledgments}
We
acknowledge support from the US NSF PHY-2207283 and the Welch Foundation (Grant No. C-1669).
\end{acknowledgments}

\appendix
\begin{widetext}
\section{Generalized Holstein-Primakoff mapping}
\label{AppA}

First, we start by expanding $\Theta$ presented in Eq.~(\ref{HHP}):
\begin{eqnarray}
    &&\Theta \approx \sqrt{N} \beta - \frac{1}{2\beta}(\beta_1 d_1^{\dagger}+\beta_1^*d_1 + \beta_2 d_2^{\dagger}+\beta_2^*d_2)
     - \frac{1}{2\beta \sqrt{N}}\left[(d_1^{\dagger}d_1 + d_2^{\dagger}d_2)+\frac{(\beta_1 d_1^{\dagger}+\beta_1^*d_1 + \beta_2 d_2^{\dagger}+\beta_2^*d_2)^2}{4 \beta^2} \right],
\end{eqnarray}
where we have kept only powers of $N$ that allow us to recast the Hamiltonian in the form of Eq.~(\ref{Eqthreeterms}). $H_0$ is explicitly given in Eq.~(\ref{H0}) and the derivatives of $H_0$ with respect to the order parameters are given by equations
\begin{eqnarray}
\frac{\partial H_0}{\partial \alpha} &=& \omega \alpha^* + g_1 \beta_1^*\beta_2 + g_2\beta_2^* \beta_1 + g_1 \gamma \beta \beta_2^* + g_2\gamma \beta \beta_2 \nonumber \\
\frac{\partial H_0}{\partial \beta_1} &=& (2-\delta)\Omega \beta_1^* + g_1 \alpha^*\beta_2^* + g_2 \alpha \beta_2^*  -\frac{g_1 \gamma \beta_1^*}{2 \beta}(\alpha \beta_2^* + \textnormal{c.c}) -\frac{g_2 \gamma \beta_1^*}{2 \beta}(\alpha \beta_2 + \textnormal{c.c}) \nonumber \\
\frac{\partial H_0}{\partial \beta_2} &=& \Omega \beta_2^* + g_1 \alpha \beta_1^* + g_2 \alpha^* \beta_1^* + g_1 \gamma \beta \alpha^* + g_2 \gamma \beta \alpha   -\frac{g_1 \gamma \beta_2^*}{2 \beta}(\alpha \beta_2^* + \textnormal{c.c}) -\frac{g_2 \gamma \beta_2^*}{2 \beta}(\alpha \beta_2 + \textnormal{c.c}) \,.
\end{eqnarray}
Additionally, $\frac{\partial H_0}{\partial \alpha^*} = \left(\frac{\partial H_0}{\partial \alpha} \right)^*$, $\frac{\partial H_0}{\partial \beta_1^*} = \left(\frac{\partial H_0}{\partial \beta_1} \right)^*$, and $\frac{\partial H_0}{\partial \beta_2^*} = \left(\frac{\partial H_0}{\partial \beta_2} \right)^*$. Clearly, minimization of $H_0$ requires that simultaneously all the six derivatives presented above are equal to zero. Expanding the terms proportional to $\sqrt{N}$ we find that:
\begin{eqnarray}
H_1 &=& \left( \frac{\partial H_0}{\partial \alpha} \right) c + \left( \frac{\partial H_0}{\partial \alpha^*} \right) c^{\dagger} + \left( \frac{\partial H_0}{\partial \beta_1} \right) d_1 + \left( \frac{\partial H_0}{\partial \beta_1^*} \right) d_1^{\dagger} + \left( \frac{\partial H_0}{\partial \beta_2} \right) d_2 + \left( \frac{\partial H_0}{\partial \beta_2^*} \right) d_2^{\dagger}  \,.
\end{eqnarray}
This means that as long as we always consider the values of the order parameters that minimize the energy then $H_1=0$.

$H_2$ can be rewritten in the general form given in Eq.~(\ref{H2}), with $\vec{v}=(c^{\dagger},d_1^{\dagger},d_1,c,d_1,d_2)$, the explicit values of the $\mathcal{C}_{jk}$ are given by:
\begin{eqnarray}
&\mathcal{C}_{11} = \mathcal{C}_{44} = 0, \quad \mathcal{C}_{14} = \mathcal{C}_{41}=\omega/2 \nonumber \\
&\mathcal{C}_{22} = \mathcal{C}_{55}^* = -\frac{\gamma  \beta_1^2}{8\beta^3}(g_1(\alpha \beta_2^*+\textnormal{c.c})+g_2(\alpha \beta_2+\textnormal{c.c})) \nonumber \\
&\mathcal{C}_{25} = \mathcal{C}_{52}^* = \frac{1}{2}\left((2-\delta)\Omega - \frac{g_1 \gamma}{2\beta}\left(1+\frac{\vert \beta_1 \vert^2}{2\beta^2}\right)(\alpha \beta_2^* + \textnormal{c.c})\right) - \frac{g_2 \gamma}{4\beta}\left(1+\frac{\vert \beta_1 \vert^2}{2\beta^2}\right)(\alpha \beta_2 + \textnormal{c.c}) \nonumber \\
 &\mathcal{C}_{33} = \mathcal{C}_{66}^* = -\frac{\gamma \beta_2^2}{8\beta^3}(g_1(\alpha \beta_2^* + \textnormal{c.c})+g_2(\alpha \beta_2 + \textnormal{c.c}))-\frac{\gamma}{2\beta}(g_1 \alpha \beta_2 + g_2 \alpha^* \beta_2) \nonumber \\
 &\mathcal{C}_{36} = \mathcal{C}_{63}^* = \frac{1}{2}\left(\Omega -\frac{g_1 \gamma}{2\beta}\left(2+\frac{\vert \beta_2\vert ^2}{2 \beta^2}\right)(\alpha \beta_2^* + \textnormal{c.c}) \right) - \frac{g_2 \gamma}{4\beta}\left(2+\frac{\vert \beta_2 \vert^2}{2\beta^2}\right)(\alpha \beta_2 + \textnormal{c.c}) \nonumber \\
 & \mathcal{C}_{12} = \mathcal{C}_{21} = \mathcal{C}_{45}^* = \mathcal{C}_{54}^* = \frac{1}{2}g_2\beta_2 -\frac{\gamma}{4\beta}(g_1 \beta_2 \beta_1 + g_2 \beta_2^* \beta_1) \nonumber  \\
 & \mathcal{C}_{15} = \mathcal{C}_{51} = \mathcal{C}_{42}^* = \mathcal{C}_{24}^* = \frac{1}{2} g_1 \beta_2^* -\frac{\gamma}{4\beta}(g_1 \beta_2 \beta_1^* + g_2 \beta_2^* \beta_1^*) \nonumber \\
 &\mathcal{C}_{13} = \mathcal{C}_{31} = \mathcal{C}_{46}^* = \mathcal{C}_{64}^* = \frac{1}{2}\left( g_1 \beta_1 - \frac{g_1 \gamma \beta_2^2}{2\beta} \right) +\frac{1}{2}\left(g_2 \gamma \beta - \frac{g_2 \gamma \vert \beta_2 \vert^2}{2\beta} \right) \nonumber \\
 &\mathcal{C}_{16} = \mathcal{C}_{61}=\mathcal{C}_{43}^* = \mathcal{C}_{34}^* = \frac{1}{2}\left( g_1 \gamma \beta - \frac{g_1 \gamma \vert \beta_2 \vert ^2}{2 \beta}\right) + \frac{1}{2} \left(g_2 \beta_1^* - \frac{g_2 \gamma (\beta_2^*)^2}{2\beta} \right) \nonumber \\
 &\mathcal{C}_{23} = \mathcal{C}_{32} = \mathcal{C}_{56}^* = \mathcal{C}_{65}^*= - \frac{g_1 \gamma \alpha \beta_1}{4\beta} - \frac{g_2 \gamma \alpha^* \beta_1}{4\beta}  -\frac{g_1\gamma \beta_2 \beta_1}{8 \beta^3}(\alpha\beta_2^* + \textnormal{c.c})  -\frac{g_2\gamma \beta_2 \beta_1}{8 \beta^3}(\alpha\beta_2 + \textnormal{c.c}) \nonumber \\
&\mathcal{C}_{26} = \mathcal{C}_{62} = \mathcal{C}_{35}^* = \mathcal{C}_{53}^* = \frac{1}{2}(g_1 \alpha + g_2\alpha^*) -\frac{g_1\gamma \alpha^* \beta_1}{4\beta} - \frac{g_2 \gamma \alpha \beta_1}{4\beta }  -\frac{g_1 \gamma \beta_1 \beta_2^*}{8\beta^3}(\alpha \beta_2^* + \textnormal{c.c}) -\frac{g_2 \gamma \beta_1 \beta_2^*}{8\beta^3}(\alpha \beta_2 + \textnormal{c.c}) \,.
\end{eqnarray}

\end{widetext}

\section{Perturbation theory}
\label{AppB}

In the SRA and SRB phases, the mean-field value of $\langle a \rangle = \alpha $ is purely real and purely imaginary, respectively. Here we explicitly compute the critical boundaries for the SRA phase, but an identical procedure follows for the SRB phase.

First, replacing $a$ and $a^\dag$ in Eq.~(\ref{TDMH}) by their expectation value $\alpha$, we obtain the mean-field Hamiltonian as 
\begin{eqnarray}
    H_{\rm MF}/N\Omega \!&=& \!\frac{1}{\lambda_+^2} \alpha^2 + (1-\delta)P_{11} - \Omega P_{33} \nonumber \\
    &&+\alpha \, ( P_{12}+\gamma  P_{23}+ P_{21}+\gamma P_{32}) \, ,
    \label{AppTDMH}
\end{eqnarray}
where we have rescaled $\frac{(g_1+g_2)\alpha}{\Omega \sqrt{N}} \rightarrow \alpha $.
Near the critical line (either a second-order phase transition or a TP), we expect $\alpha$ to be very small. We can then apply the time-independent perturbation theory, treating the first line of Eq.~(\ref{AppTDMH}) as the unperturbed Hamiltonian and the second line the perturbing Hamiltonian. 

If we keep the perturbation expansion up to the sixth order, the mean-field energy will have the form
\begin{equation}
    E_{\rm MF}/N \Omega = p_0 + p_1 \alpha^2 + p_2 \alpha^4 + p_3 \alpha^6\,,
\end{equation}
where the $p_i$ coefficients are explicitly given as
\begin{eqnarray}
    &p_0 = -1, \quad p_1 = \frac{1}{\lambda_+^2}-\gamma^2, \quad p_2 = \gamma^2\left(\gamma^2 - \frac{1}{2-\delta}\right) \,, \nonumber \\
    &p_3 = \gamma^2 \left(-2 \gamma^4 + \frac{3 \gamma^2}{2-\delta}+\frac{\gamma^2}{(2-\delta)^2}-\frac{1}{(2-\delta)^2}\right) \, .
\end{eqnarray}
Using the standard Landau theory analysis \cite{chaikin_lubensky_1995}, if $p_2>0$, the line $p_1=0$ represents a second-order boundary which leads to
\[ \gamma^2=\frac{1}{\lambda_+^2}>\frac{1}{2-\delta}\]and the TP is determined by the conditions $p_1=p_2=0$ and $p_3>0$, i.e., \[ \gamma^2=\frac{1}{\lambda_+^2}=\frac{1}{2-\delta}\]
These are the results reported in Eq.~(\ref{TPcond1}) in the main text.

An alternative and straightforward method to find these two conditions was described in \cite{Youjiang2021} for tridiagonal Hamiltonians like our TDM Hamiltonian. In order to use that result, we rewrite the Hamiltonian in the consistent notation
\begin{equation}
    H_{\rm MF}/N \Omega =  \left(\frac{1}{\lambda_+^2} \alpha^2 + 1 \right) \mathbb{I} + \alpha d + h \,,
\end{equation}
here  $\mathbb{I}$ is the 3 by 3 identity matrix, and the $d$ and $h$ matrices are defined as 
\begin{equation}
        d = \begin{pmatrix}
        0&1&0\\1&0&\gamma\\0&\gamma&0
    \end{pmatrix}, \quad h= \begin{pmatrix}
        (2-\delta)&0&0\\0&1 &0\\0&0&0
    \end{pmatrix} \,.
\end{equation}
In this notation, the critical conditions are given by
\begin{equation}
    \vert d_{k,k-1}\vert^2 = \frac{1}{\lambda_+^2} h_{k,k}, \quad \text{for } k=2,3 \,.
\end{equation}
This yields the two critical equations $\gamma^2 = 1/\lambda_+^2$ and $\lambda_+^2 = 2-\delta$ as presented in Eq.~(\ref{TPcond1}). These two constraints are equivalent to $p_1=0$ and $p_2=0$, respectively. A similar procedure but using $\langle a \rangle = i \alpha$, leads to Eq.~(\ref{TPcond2}) for the SRB phase.

\section{Bogoliubov transformation}
\label{AppC}

Since the Hamiltonian in Eq.~(\ref{H2}) is bilinear in the annihilation and creation operators we can diagonalize it by doing a Bogoliubov transformation.

First, we can rewrite $H_2$ as:
\begin{equation}
    H_2 = \vec{v} \mathcal{M} \vec{v}^{\dagger}\,,
\end{equation}
where, in our current notation, $\mathcal{M}$ is given by:
\begin{equation}
\mathcal{M}=
\begin{pmatrix}
\mathcal{C}_{14}& \mathcal{C}_{15}&\mathcal{C}_{16} & \mathcal{C}_{11} & \mathcal{C}_{12} & \mathcal{C}_{13} \\
\mathcal{C}_{24}& \mathcal{C}_{25}&\mathcal{C}_{26} & \mathcal{C}_{21} & \mathcal{C}_{22} & \mathcal{C}_{23} \\
\mathcal{C}_{34}& \mathcal{C}_{35}&\mathcal{C}_{36} & \mathcal{C}_{31} & \mathcal{C}_{32} & \mathcal{C}_{33} \\
\mathcal{C}_{44}& \mathcal{C}_{45}&\mathcal{C}_{46} & \mathcal{C}_{41} & \mathcal{C}_{42} & \mathcal{C}_{43} \\
\mathcal{C}_{54}& \mathcal{C}_{55}&\mathcal{C}_{56} & \mathcal{C}_{51} & \mathcal{C}_{52} & \mathcal{C}_{53} \\
\mathcal{C}_{64}& \mathcal{C}_{65}&\mathcal{C}_{66} & \mathcal{C}_{61} & \mathcal{C}_{62} & \mathcal{C}_{63} \\
\end{pmatrix} \, .
\end{equation}
Now, let us consider a Bogoliubov transformation $T$ such that $\vec{v}^{\dagger} = T \vec{u}^{\dagger}$, where $\vec{u} = (a_1^{\dagger},a_2^{\dagger},a_3^{\dagger},a_1,a_2,a_3)$ are a new set of annihilation and creation operators. Our objective is to find the transformation $T$ such that $H_2$ is diagonalized as in Eq.~(\ref{H2bogo}). 

Since we require the operators in $\vec{u}$ to follow canonical bosonic commutation relations, namely, $[a_j,a_k^{\dagger}] = \delta_{jk}$, $[a_j,a_k]=0$, and $[a_j^{\dagger},a_k^{\dagger}]=0$, it follows that $T$ is constrained by
\begin{equation}
    T^{\dagger} \Gamma T = \Gamma \,,
\end{equation}
where $\Gamma$ is a diagonal matrix with diagonal given by $(1,1,1,-1,-1,-1)$. Since we are looking for $T$ such that $T^{\dagger} \mathcal{M} T$ is diagonal with two-fold degenerate eigenvalues $\varepsilon_1$, $\varepsilon_2$, and $\varepsilon_3$, then it follows that $T^{\dagger} \Gamma^ 2\mathcal{M} T = \Gamma T^{\dagger} = \Gamma T^{-1}\Gamma \mathcal{M} T$, which means that $T^{-1}\Gamma \mathcal{M} T$ is a diagonal matrix with eigenvalues $\varepsilon_1$, $\varepsilon_2$, $\varepsilon_3$, $-\varepsilon_1$, $-\varepsilon_2$, and $-\varepsilon_3$. Then, by simply diagonalizing the matrix $\Gamma \mathcal{M}$ we can find both the transformation matrix $T$ as well as the corresponding eigenvalues. 

\section{Gell-Mann matrices}
\label{AppD}

The Gell-Mann matrices are a group of eight 3 by 3 matrices that generate the $SU(3)$ algebra. They are explicitly defined as \cite{gellmann}:
\begin{eqnarray}
    &\Lambda_1^{(k)} = \begin{pmatrix}
        0&1&0\\1&0&0\\0&0&0
    \end{pmatrix}, \quad \Lambda_2^{(k)} = \begin{pmatrix}
        0&-i&0\\i&0&0\\0&0&0
    \end{pmatrix} \nonumber \\
    &\Lambda_3^{(k)} = \begin{pmatrix}
        1&0&0\\0&-1&0\\0&0&0
    \end{pmatrix}, \quad \Lambda_4^{(k)} = \begin{pmatrix}
        0&0&1\\0&0&0\\1&0&0
    \end{pmatrix} \nonumber \\
    &\Lambda_5^{(k)} = \begin{pmatrix}
        0&0&-i\\0&0&0\\i&0&0
    \end{pmatrix}, \quad \Lambda_6^{(k)} = \begin{pmatrix}
        0&0&0\\0&0&1\\0&1&0
    \end{pmatrix} \nonumber \\
    &\Lambda_7^{(k)} = \begin{pmatrix}
        0&0&0\\0&0&-i\\0&i&0
    \end{pmatrix}, \quad \Lambda_8^{(k)} = \frac{1}{\sqrt{3}}\begin{pmatrix}
        1&0&0\\0&1&0\\0&0&-2
    \end{pmatrix} \, .
\end{eqnarray}
where the superscript $k$ indicates that these are single-particle operators associated with the $k$th atom. The commutation and anticommutation relations of the Gell-Mann matrices are given, respectively, by:
\begin{eqnarray}
    [\Lambda_j^{(n)}, \Lambda_k^{(n)}] &=& 2i\sum_lf_{jkl}\Lambda_l^{(n)} \,,
\\
    \{\Lambda_j^{(n)}, \Lambda_k^{(n)}\} &=& \frac{4}{3}\delta_{jk} \mathbb{I} + 2\sum_ld_{jkl} \Lambda_l^{(n)} \,.
\end{eqnarray}
Here $f_{jkl}$ are totally antisymmetric structure constants and most of them vanish, for a list of the nonzero values of $f_{jkl}$ see Ref.~\cite{greiner2012quantum}. The $d_{jkl}$ are totally symmetric constants defined explicitly by $d_{jkl}=\frac{1}{4}\text{tr}(\{\Lambda_j, \Lambda_k\}\Lambda_l)$. These $d_{jkl}$ constants are also used to define one of the Casimir operators, see Eq.~(\ref{casimir}). By construction, the collective operators $\Lambda_j = \sum_{k=1}^N \Lambda_j^{(k)}$ used in the main text follow the same commutation/anticommutation relations given above.

\section{Matrix Elements of $SU(3)$ Dicke states}
\label{AppE}

Here we list how each operator acts on the generalized Dicke states $\vert t, t_z \rangle$. Both $T_z$ and $Y$ are diagonal in this basis
\begin{equation}
    T_z \vert t, t_z \rangle = t_z \vert t,t_z \rangle, \quad Y \vert t, t_z \rangle = \left( 2t - \frac{2N}{3}\right) \vert t, t_z \rangle \, ,
\end{equation}
where $N$ is the number of atoms. In the second relation, we have used the fact that $y$ the eigenvalue of $Y$ is not independent of the eigenvalue $t$ in a totally symmetric representation. 

Since $T_\pm$ and $T_z$ define an $SU(2)$ subalgebra, the matrix elements of $T_{\pm}$ are defined as
\begin{equation}
T_{\pm} \vert t,t_z\rangle= \sqrt{t(t+1)-t_z(t_z \pm 1)} \,\vert t,t_z\pm 1 \rangle \, .
\end{equation}

Finally, for a totally symmetric representation $D(N,0)$, the matrix elements of the ladder operators $U_{\pm}$ are given by \cite{deSwart}
\begin{eqnarray}
    U_+ \vert t,t_z \rangle = \sqrt{(t-t_z+1)(N-2t)} \,\vert t+1/2,t_z-1/2 \rangle \,,\nonumber \\
    U_- \vert t,t_z \rangle = \sqrt{(t-t_z)(N-2t+1)} \,\vert t-1/2,t_z+1/2  \rangle \,.\nonumber
\end{eqnarray}

With all these matrix elements being defined, we can construct a matrix representation of Hamiltonian~(\ref{TDMHCW}) and perform exact diagonalization.

\section{Potential experimental realization}
\label{AppF}

\begin{figure}[t!]
\includegraphics[width=0.48\textwidth]{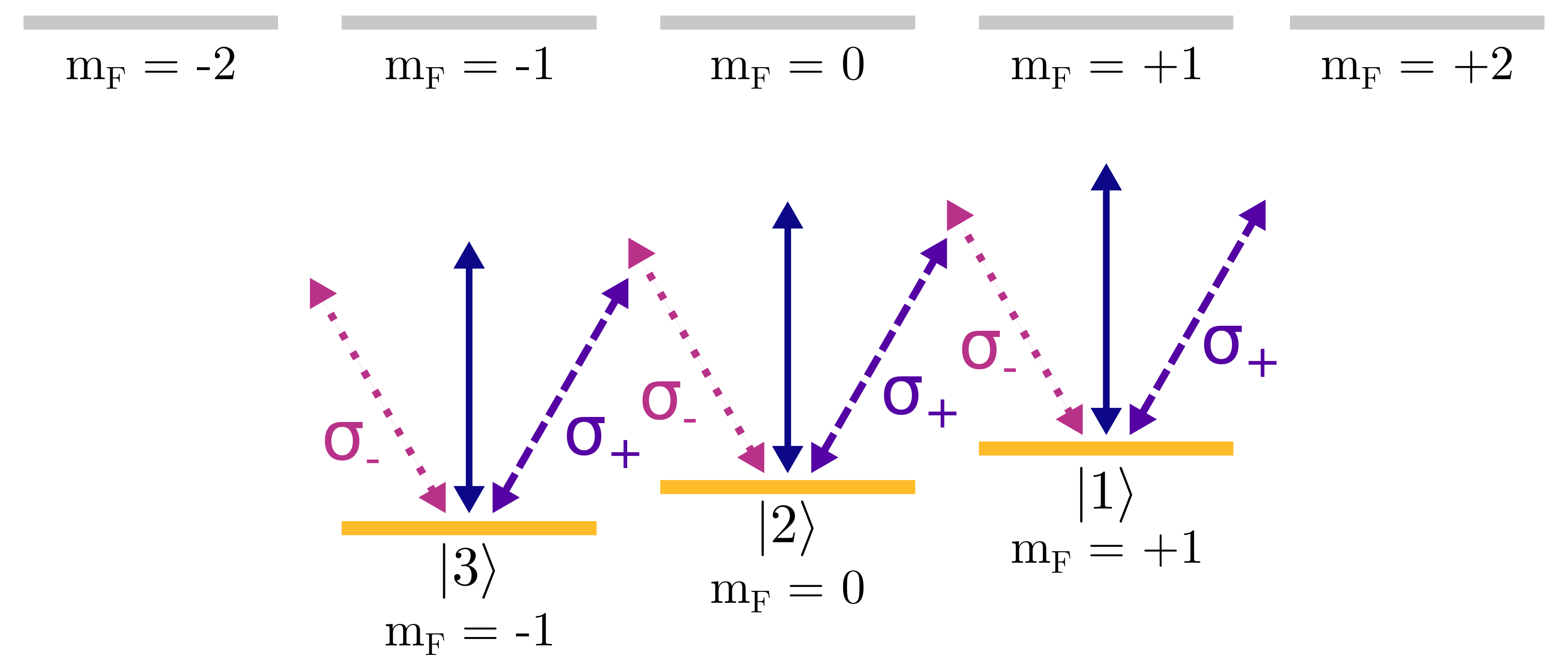}
\caption{Schematics of the proposed experimental implementation~\cite{Masson2017}. Three hyperfine states, for example, those in the $F=1$ manifold in $^{87}$Rb, are coupled to a manifold of excited states through a cavity mode (blue solid arrows) and through circularly polarized lasers with polarization $\sigma_+$ (purple dashed arrows) and $\sigma_-$ (pink dotted arrows). By highly detuning the excited states the system becomes effectively a three-leveled one, with states $\vert 1 \rangle$, $\vert 2 \rangle$ and $\vert 3 \rangle$ coupled to each other.}
\label{FigAppendix}
\end{figure}

We consider three hyperfine ground levels $\vert 1 \rangle$, $\vert 2 \rangle$, $\vert 3 \rangle$ that could be the ones on the $F=1$ ground state of $^{87}$Rb as depicted in Fig.~\ref{FigAppendix}. These states are coupled to a manifold of excited states, with $F=2$ in this case, through a cavity mode with $\pi$ polarization, and lasers with circular $\sigma_{\pm}$ polarizations. Importantly enough, the presence of both $\sigma_+$ and $\sigma_-$ lasers allows for tuning independently the co- and counter-rotating terms ($g_1$ and $g_2$ in this work) by modifying each laser's parameters. Additionally, the bare energies of each ground state can be modified through Zeeman shifts using magnetic fields or using microwaves through the AC Stark shift~\cite{PhysRevA.73.041602,PhysRevA.79.023406}.
 
We consider that the one-photon transitions between the ground and the excited levels are far off resonance such that the excited levels can be adiabatically eliminated,
making the system effectively three-leveled, with the three levels now coupled through the cavity and laser fields. We refer to~\cite{Masson2017}, where this experimental realization was originally proposed, for a complete derivation of the Hamiltonian parameters in terms of the laser parameters and detunings.

Finally, another important consequence of the chosen setup is that since the states $\vert 1 \rangle$, $\vert 2 \rangle$ and $\vert 3 \rangle$ are chosen to be ground states (do not have any spontaneous decay channels), we can omit any individual and collective atomic decoherence channels and focus only in the photons leaking out of the cavity, as in Eq.~(\ref{AdjLindblad}).

\section{Stability of normal phases}
\label{AppG}

In order to find the stability boundaries of the normal phases we can use the $H_2$ Hamiltonian in Eq.~(\ref{H2}). For the NP3, $H_2$ is simply given by
\begin{eqnarray}
    H_2 &=& \omega c^{\dagger}c + 2\Omega(1-\delta) d_1^{\dagger}d_1 + \Omega d_2^{\dagger}d_2 \nonumber \\
   & &+g_2 \gamma (c^{\dagger}d_2^{\dagger}+cd_2) + g_1 \gamma (c^{\dagger}d_2 + cd_2^{\dagger})\,.
    \label{AppH2}
\end{eqnarray}
Now, we can compute the Lindblad Eq.~(\ref{AdjLindblad}) for $c$, $d_1$, and $d_2$ using $H_2$ as a Hamiltonian. Note that since we are in a normal phase, here $c=a$. The six resulting equations can be written in matrix form as
\begin{eqnarray}
\frac{d\vec{x}}{dt}= 
\begin{pmatrix}
-\kappa& \omega& 0& 0& 0& 0\\
-\omega& -\kappa&0&0&-2 g \gamma&0\\
0&0&0&2\Omega&0&0\\
0&0&-2\Omega&0&0&0\\
0&0&0&0&0&\Omega \\
-2 g \gamma & 0 & 0 &0 &-\Omega & 0
\end{pmatrix}
\vec{x} \,,
\end{eqnarray}
where we have set $g_1=g_2=g$ and $\delta=0$ in order to be consistent with the main text discussion. Here $\vec{x}^T = (\text{Re}(\langle c \rangle),\text{Im}(\langle c \rangle),\text{Re}(\langle d_1 \rangle),\text{Im}(\langle d_1 \rangle),\text{Re}(\langle d_2 \rangle),\text{Im}(\langle d_2 \rangle))$. The eigenvalues of the matrix above determine whether the NP3 phase represents a stable steady state or not. If the real part of all eigenvalues is negative the NP3 phase is stable; on the other hand, if at least one of the eigenvalues has a positive real part, the phase is unstable.

Since we are interested in the boundary where the phase becomes unstable, it is important to determine when the matrix develops a zero eigenvalue. Then, we set the determinant of the matrix equation to zero, leading to:
\begin{equation}
    -4\Omega^3(4g^2\gamma^2\omega - \kappa^2 \Omega - \omega^2 \Omega)=0\,.
\end{equation}
After some algebra, and since $\lambda_+ = \lambda_1+\lambda_2 = 2g/\sqrt{\omega \Omega}$ in this case, we obtain the boundary equation
\begin{equation}
    \lambda_+^2 = \frac{1+\frac{\kappa^2}{\omega^2}}{\gamma^2}\,,
\end{equation}
which is given as the first equation in Eq.~(\ref{stabilitybounds}). As explained in Ref.~\cite{Hayn2011}, if we want to consider the case for the NP2 phase, it is required to repeat the process of the generalized Holstein-Primakoff mapping but using $\vert 2 \rangle$ as the reference state. Once the mapping is done, one can take the $H_2$ Hamiltonian found for the NP2 phase and compute the Lindblad equations. Performing a similar stability analysis the stability boundary of the NP2 phase is found to be
\begin{equation}
    \lambda_+^2 = \frac{1+\frac{\kappa^2}{\omega^2}}{1-\gamma^2}\,.
\end{equation}
which is given as the second equation in Eq.~(\ref{stabilitybounds}).

\bibliography{refs}{}

\begin{thebibliography}{62}%
\makeatletter
\providecommand \@ifxundefined [1]{%
 \@ifx{#1\undefined}
}%
\providecommand \@ifnum [1]{%
 \ifnum #1\expandafter \@firstoftwo
 \else \expandafter \@secondoftwo
 \fi
}%
\providecommand \@ifx [1]{%
 \ifx #1\expandafter \@firstoftwo
 \else \expandafter \@secondoftwo
 \fi
}%
\providecommand \natexlab [1]{#1}%
\providecommand \enquote  [1]{``#1''}%
\providecommand \bibnamefont  [1]{#1}%
\providecommand \bibfnamefont [1]{#1}%
\providecommand \citenamefont [1]{#1}%
\providecommand \href@noop [0]{\@secondoftwo}%
\providecommand \href [0]{\begingroup \@sanitize@url \@href}%
\providecommand \@href[1]{\@@startlink{#1}\@@href}%
\providecommand \@@href[1]{\endgroup#1\@@endlink}%
\providecommand \@sanitize@url [0]{\catcode `\\12\catcode `\$12\catcode `\&12\catcode `\#12\catcode `\^12\catcode `\_12\catcode `\%12\relax}%
\providecommand \@@startlink[1]{}%
\providecommand \@@endlink[0]{}%
\providecommand \url  [0]{\begingroup\@sanitize@url \@url }%
\providecommand \@url [1]{\endgroup\@href {#1}{\urlprefix }}%
\providecommand \urlprefix  [0]{URL }%
\providecommand \Eprint [0]{\href }%
\providecommand \doibase [0]{https://doi.org/}%
\providecommand \selectlanguage [0]{\@gobble}%
\providecommand \bibinfo  [0]{\@secondoftwo}%
\providecommand \bibfield  [0]{\@secondoftwo}%
\providecommand \translation [1]{[#1]}%
\providecommand \BibitemOpen [0]{}%
\providecommand \bibitemStop [0]{}%
\providecommand \bibitemNoStop [0]{.\EOS\space}%
\providecommand \EOS [0]{\spacefactor3000\relax}%
\providecommand \BibitemShut  [1]{\csname bibitem#1\endcsname}%
\let\auto@bib@innerbib\@empty
\bibitem [{\citenamefont {Dicke}(1954)}]{Dicke1954}%
  \BibitemOpen
  \bibfield  {author} {\bibinfo {author} {\bibfnamefont {R.~H.}\ \bibnamefont {Dicke}},\ }\href {https://doi.org/10.1103/PhysRev.93.99} {\bibfield  {journal} {\bibinfo  {journal} {Phys. Rev.}\ }\textbf {\bibinfo {volume} {93}},\ \bibinfo {pages} {99} (\bibinfo {year} {1954})}\BibitemShut {NoStop}%
\bibitem [{\citenamefont {Gross}\ \emph {et~al.}(1976)\citenamefont {Gross}, \citenamefont {Fabre}, \citenamefont {Pillet},\ and\ \citenamefont {Haroche}}]{gross1976}%
  \BibitemOpen
  \bibfield  {author} {\bibinfo {author} {\bibfnamefont {M.}~\bibnamefont {Gross}}, \bibinfo {author} {\bibfnamefont {C.}~\bibnamefont {Fabre}}, \bibinfo {author} {\bibfnamefont {P.}~\bibnamefont {Pillet}},\ and\ \bibinfo {author} {\bibfnamefont {S.}~\bibnamefont {Haroche}},\ }\href {https://doi.org/10.1103/PhysRevLett.36.1035} {\bibfield  {journal} {\bibinfo  {journal} {Phys. Rev. Lett.}\ }\textbf {\bibinfo {volume} {36}},\ \bibinfo {pages} {1035} (\bibinfo {year} {1976})}\BibitemShut {NoStop}%
\bibitem [{\citenamefont {Scheibner}\ \emph {et~al.}(2007)\citenamefont {Scheibner}, \citenamefont {Schmidt}, \citenamefont {Worschech}, \citenamefont {Forchel}, \citenamefont {Bacher}, \citenamefont {Passow},\ and\ \citenamefont {Hommel}}]{scheibner2007superradiance}%
  \BibitemOpen
  \bibfield  {author} {\bibinfo {author} {\bibfnamefont {M.}~\bibnamefont {Scheibner}}, \bibinfo {author} {\bibfnamefont {T.}~\bibnamefont {Schmidt}}, \bibinfo {author} {\bibfnamefont {L.}~\bibnamefont {Worschech}}, \bibinfo {author} {\bibfnamefont {A.}~\bibnamefont {Forchel}}, \bibinfo {author} {\bibfnamefont {G.}~\bibnamefont {Bacher}}, \bibinfo {author} {\bibfnamefont {T.}~\bibnamefont {Passow}},\ and\ \bibinfo {author} {\bibfnamefont {D.}~\bibnamefont {Hommel}},\ }\href {https://doi.org/10.1038/nphys494} {\bibfield  {journal} {\bibinfo  {journal} {Nat. Phys.}\ }\textbf {\bibinfo {volume} {3}},\ \bibinfo {pages} {106} (\bibinfo {year} {2007})}\BibitemShut {NoStop}%
\bibitem [{\citenamefont {Skribanowitz}\ \emph {et~al.}(1973)\citenamefont {Skribanowitz}, \citenamefont {Herman}, \citenamefont {MacGillivray},\ and\ \citenamefont {Feld}}]{skribanowitz}%
  \BibitemOpen
  \bibfield  {author} {\bibinfo {author} {\bibfnamefont {N.}~\bibnamefont {Skribanowitz}}, \bibinfo {author} {\bibfnamefont {I.~P.}\ \bibnamefont {Herman}}, \bibinfo {author} {\bibfnamefont {J.~C.}\ \bibnamefont {MacGillivray}},\ and\ \bibinfo {author} {\bibfnamefont {M.~S.}\ \bibnamefont {Feld}},\ }\href {https://doi.org/10.1103/PhysRevLett.30.309} {\bibfield  {journal} {\bibinfo  {journal} {Phys. Rev. Lett.}\ }\textbf {\bibinfo {volume} {30}},\ \bibinfo {pages} {309} (\bibinfo {year} {1973})}\BibitemShut {NoStop}%
\bibitem [{\citenamefont {Hepp}\ and\ \citenamefont {Lieb}(1973)}]{hepplieb}%
  \BibitemOpen
  \bibfield  {author} {\bibinfo {author} {\bibfnamefont {K.}~\bibnamefont {Hepp}}\ and\ \bibinfo {author} {\bibfnamefont {E.~H.}\ \bibnamefont {Lieb}},\ }\href {https://doi.org/10.1103/PhysRevA.8.2517} {\bibfield  {journal} {\bibinfo  {journal} {Phys. Rev. A}\ }\textbf {\bibinfo {volume} {8}},\ \bibinfo {pages} {2517} (\bibinfo {year} {1973})}\BibitemShut {NoStop}%
\bibitem [{\citenamefont {Wang}\ and\ \citenamefont {Hioe}(1973)}]{hioe}%
  \BibitemOpen
  \bibfield  {author} {\bibinfo {author} {\bibfnamefont {Y.~K.}\ \bibnamefont {Wang}}\ and\ \bibinfo {author} {\bibfnamefont {F.~T.}\ \bibnamefont {Hioe}},\ }\href {https://doi.org/10.1103/PhysRevA.7.831} {\bibfield  {journal} {\bibinfo  {journal} {Phys. Rev. A}\ }\textbf {\bibinfo {volume} {7}},\ \bibinfo {pages} {831} (\bibinfo {year} {1973})}\BibitemShut {NoStop}%
\bibitem [{\citenamefont {Masson}\ and\ \citenamefont {Parkins}(2019)}]{masson2019}%
  \BibitemOpen
  \bibfield  {author} {\bibinfo {author} {\bibfnamefont {S.~J.}\ \bibnamefont {Masson}}\ and\ \bibinfo {author} {\bibfnamefont {S.}~\bibnamefont {Parkins}},\ }\href {https://doi.org/10.1103/PhysRevA.99.023822} {\bibfield  {journal} {\bibinfo  {journal} {Phys. Rev. A}\ }\textbf {\bibinfo {volume} {99}},\ \bibinfo {pages} {023822} (\bibinfo {year} {2019})}\BibitemShut {NoStop}%
\bibitem [{\citenamefont {Hayashida}\ \emph {et~al.}(2023)\citenamefont {Hayashida}, \citenamefont {Makihara}, \citenamefont {Marquez~Peraca}, \citenamefont {Fallas~Padilla}, \citenamefont {Pu}, \citenamefont {Kono},\ and\ \citenamefont {Bamba}}]{hayashida2023perfect}%
  \BibitemOpen
  \bibfield  {author} {\bibinfo {author} {\bibfnamefont {K.}~\bibnamefont {Hayashida}}, \bibinfo {author} {\bibfnamefont {T.}~\bibnamefont {Makihara}}, \bibinfo {author} {\bibfnamefont {N.}~\bibnamefont {Marquez~Peraca}}, \bibinfo {author} {\bibfnamefont {D.}~\bibnamefont {Fallas~Padilla}}, \bibinfo {author} {\bibfnamefont {H.}~\bibnamefont {Pu}}, \bibinfo {author} {\bibfnamefont {J.}~\bibnamefont {Kono}},\ and\ \bibinfo {author} {\bibfnamefont {M.}~\bibnamefont {Bamba}},\ }\href {https://doi.org/10.1038/s41598-023-29202-x} {\bibfield  {journal} {\bibinfo  {journal} {Sci. Rep.}\ }\textbf {\bibinfo {volume} {13}},\ \bibinfo {pages} {2526} (\bibinfo {year} {2023})}\BibitemShut {NoStop}%
\bibitem [{\citenamefont {Emary}\ and\ \citenamefont {Brandes}(2003)}]{Emary2003}%
  \BibitemOpen
  \bibfield  {author} {\bibinfo {author} {\bibfnamefont {C.}~\bibnamefont {Emary}}\ and\ \bibinfo {author} {\bibfnamefont {T.}~\bibnamefont {Brandes}},\ }\href {https://doi.org/10.1103/PhysRevE.67.066203} {\bibfield  {journal} {\bibinfo  {journal} {Phys. Rev. E}\ }\textbf {\bibinfo {volume} {67}},\ \bibinfo {pages} {066203} (\bibinfo {year} {2003})}\BibitemShut {NoStop}%
\bibitem [{\citenamefont {Xu}\ and\ \citenamefont {Pu}(2019)}]{Xu1}%
  \BibitemOpen
  \bibfield  {author} {\bibinfo {author} {\bibfnamefont {Y.}~\bibnamefont {Xu}}\ and\ \bibinfo {author} {\bibfnamefont {H.}~\bibnamefont {Pu}},\ }\href {https://doi.org/10.1103/PhysRevLett.122.193201} {\bibfield  {journal} {\bibinfo  {journal} {Phys. Rev. Lett.}\ }\textbf {\bibinfo {volume} {122}},\ \bibinfo {pages} {193201} (\bibinfo {year} {2019})}\BibitemShut {NoStop}%
\bibitem [{\citenamefont {Skulte}\ \emph {et~al.}(2021)\citenamefont {Skulte}, \citenamefont {Kongkhambut}, \citenamefont {Ke\ss{}ler}, \citenamefont {Hemmerich}, \citenamefont {Mathey},\ and\ \citenamefont {Cosme}}]{skulte}%
  \BibitemOpen
  \bibfield  {author} {\bibinfo {author} {\bibfnamefont {J.}~\bibnamefont {Skulte}}, \bibinfo {author} {\bibfnamefont {P.}~\bibnamefont {Kongkhambut}}, \bibinfo {author} {\bibfnamefont {H.}~\bibnamefont {Ke\ss{}ler}}, \bibinfo {author} {\bibfnamefont {A.}~\bibnamefont {Hemmerich}}, \bibinfo {author} {\bibfnamefont {L.}~\bibnamefont {Mathey}},\ and\ \bibinfo {author} {\bibfnamefont {J.~G.}\ \bibnamefont {Cosme}},\ }\href {https://doi.org/10.1103/PhysRevA.104.063705} {\bibfield  {journal} {\bibinfo  {journal} {Phys. Rev. A}\ }\textbf {\bibinfo {volume} {104}},\ \bibinfo {pages} {063705} (\bibinfo {year} {2021})}\BibitemShut {NoStop}%
\bibitem [{\citenamefont {Hayn}\ and\ \citenamefont {Brandes}(2017)}]{haynthermo}%
  \BibitemOpen
  \bibfield  {author} {\bibinfo {author} {\bibfnamefont {M.}~\bibnamefont {Hayn}}\ and\ \bibinfo {author} {\bibfnamefont {T.}~\bibnamefont {Brandes}},\ }\href {https://doi.org/10.1103/PhysRevE.95.012153} {\bibfield  {journal} {\bibinfo  {journal} {Phys. Rev. E}\ }\textbf {\bibinfo {volume} {95}},\ \bibinfo {pages} {012153} (\bibinfo {year} {2017})}\BibitemShut {NoStop}%
\bibitem [{\citenamefont {Fan}\ and\ \citenamefont {Jia}(2023)}]{fan2023}%
  \BibitemOpen
  \bibfield  {author} {\bibinfo {author} {\bibfnamefont {J.}~\bibnamefont {Fan}}\ and\ \bibinfo {author} {\bibfnamefont {S.}~\bibnamefont {Jia}},\ }\href {https://doi.org/10.1103/PhysRevA.107.033711} {\bibfield  {journal} {\bibinfo  {journal} {Phys. Rev. A}\ }\textbf {\bibinfo {volume} {107}},\ \bibinfo {pages} {033711} (\bibinfo {year} {2023})}\BibitemShut {NoStop}%
\bibitem [{\citenamefont {Hayn}\ \emph {et~al.}(2011)\citenamefont {Hayn}, \citenamefont {Emary},\ and\ \citenamefont {Brandes}}]{Hayn2011}%
  \BibitemOpen
  \bibfield  {author} {\bibinfo {author} {\bibfnamefont {M.}~\bibnamefont {Hayn}}, \bibinfo {author} {\bibfnamefont {C.}~\bibnamefont {Emary}},\ and\ \bibinfo {author} {\bibfnamefont {T.}~\bibnamefont {Brandes}},\ }\href {https://doi.org/10.1103/PhysRevA.84.053856} {\bibfield  {journal} {\bibinfo  {journal} {Phys. Rev. A}\ }\textbf {\bibinfo {volume} {84}},\ \bibinfo {pages} {053856} (\bibinfo {year} {2011})}\BibitemShut {NoStop}%
\bibitem [{\citenamefont {Valencia-Tortora}\ \emph {et~al.}(2023)\citenamefont {Valencia-Tortora}, \citenamefont {Kelly}, \citenamefont {Donner}, \citenamefont {Morigi}, \citenamefont {Fazio},\ and\ \citenamefont {Marino}}]{Valencia2023}%
  \BibitemOpen
  \bibfield  {author} {\bibinfo {author} {\bibfnamefont {R.~J.}\ \bibnamefont {Valencia-Tortora}}, \bibinfo {author} {\bibfnamefont {S.~P.}\ \bibnamefont {Kelly}}, \bibinfo {author} {\bibfnamefont {T.}~\bibnamefont {Donner}}, \bibinfo {author} {\bibfnamefont {G.}~\bibnamefont {Morigi}}, \bibinfo {author} {\bibfnamefont {R.}~\bibnamefont {Fazio}},\ and\ \bibinfo {author} {\bibfnamefont {J.}~\bibnamefont {Marino}},\ }\href {https://doi.org/10.1103/PhysRevResearch.5.023112} {\bibfield  {journal} {\bibinfo  {journal} {Phys. Rev. Res.}\ }\textbf {\bibinfo {volume} {5}},\ \bibinfo {pages} {023112} (\bibinfo {year} {2023})}\BibitemShut {NoStop}%
\bibitem [{\citenamefont {Xu}\ \emph {et~al.}(2021)\citenamefont {Xu}, \citenamefont {Fallas~Padilla},\ and\ \citenamefont {Pu}}]{Youjiang2021}%
  \BibitemOpen
  \bibfield  {author} {\bibinfo {author} {\bibfnamefont {Y.}~\bibnamefont {Xu}}, \bibinfo {author} {\bibfnamefont {D.}~\bibnamefont {Fallas~Padilla}},\ and\ \bibinfo {author} {\bibfnamefont {H.}~\bibnamefont {Pu}},\ }\href {https://doi.org/10.1103/PhysRevA.104.043708} {\bibfield  {journal} {\bibinfo  {journal} {Phys. Rev. A}\ }\textbf {\bibinfo {volume} {104}},\ \bibinfo {pages} {043708} (\bibinfo {year} {2021})}\BibitemShut {NoStop}%
\bibitem [{\citenamefont {Kaluarachchi}\ \emph {et~al.}(2018)\citenamefont {Kaluarachchi}, \citenamefont {Taufour}, \citenamefont {Bud'ko},\ and\ \citenamefont {Canfield}}]{kaluarachchi}%
  \BibitemOpen
  \bibfield  {author} {\bibinfo {author} {\bibfnamefont {U.~S.}\ \bibnamefont {Kaluarachchi}}, \bibinfo {author} {\bibfnamefont {V.}~\bibnamefont {Taufour}}, \bibinfo {author} {\bibfnamefont {S.~L.}\ \bibnamefont {Bud'ko}},\ and\ \bibinfo {author} {\bibfnamefont {P.~C.}\ \bibnamefont {Canfield}},\ }\href {https://doi.org/10.1103/PhysRevB.97.045139} {\bibfield  {journal} {\bibinfo  {journal} {Phys. Rev. B}\ }\textbf {\bibinfo {volume} {97}},\ \bibinfo {pages} {045139} (\bibinfo {year} {2018})}\BibitemShut {NoStop}%
\bibitem [{\citenamefont {Kim}\ \emph {et~al.}(2002)\citenamefont {Kim}, \citenamefont {Revaz}, \citenamefont {Zink}, \citenamefont {Hellman}, \citenamefont {Rhyne},\ and\ \citenamefont {Mitchell}}]{Kim2002}%
  \BibitemOpen
  \bibfield  {author} {\bibinfo {author} {\bibfnamefont {D.}~\bibnamefont {Kim}}, \bibinfo {author} {\bibfnamefont {B.}~\bibnamefont {Revaz}}, \bibinfo {author} {\bibfnamefont {B.~L.}\ \bibnamefont {Zink}}, \bibinfo {author} {\bibfnamefont {F.}~\bibnamefont {Hellman}}, \bibinfo {author} {\bibfnamefont {J.~J.}\ \bibnamefont {Rhyne}},\ and\ \bibinfo {author} {\bibfnamefont {J.~F.}\ \bibnamefont {Mitchell}},\ }\href {https://doi.org/10.1103/PhysRevLett.89.227202} {\bibfield  {journal} {\bibinfo  {journal} {Phys. Rev. Lett.}\ }\textbf {\bibinfo {volume} {89}},\ \bibinfo {pages} {227202} (\bibinfo {year} {2002})}\BibitemShut {NoStop}%
\bibitem [{\citenamefont {Khasanov}\ \emph {et~al.}(2018)\citenamefont {Khasanov}, \citenamefont {Fernandes}, \citenamefont {Simutis}, \citenamefont {Guguchia}, \citenamefont {Amato}, \citenamefont {Luetkens}, \citenamefont {Morenzoni}, \citenamefont {Dong}, \citenamefont {Zhou},\ and\ \citenamefont {Zhao}}]{kashanov2018}%
  \BibitemOpen
  \bibfield  {author} {\bibinfo {author} {\bibfnamefont {R.}~\bibnamefont {Khasanov}}, \bibinfo {author} {\bibfnamefont {R.~M.}\ \bibnamefont {Fernandes}}, \bibinfo {author} {\bibfnamefont {G.}~\bibnamefont {Simutis}}, \bibinfo {author} {\bibfnamefont {Z.}~\bibnamefont {Guguchia}}, \bibinfo {author} {\bibfnamefont {A.}~\bibnamefont {Amato}}, \bibinfo {author} {\bibfnamefont {H.}~\bibnamefont {Luetkens}}, \bibinfo {author} {\bibfnamefont {E.}~\bibnamefont {Morenzoni}}, \bibinfo {author} {\bibfnamefont {X.}~\bibnamefont {Dong}}, \bibinfo {author} {\bibfnamefont {F.}~\bibnamefont {Zhou}},\ and\ \bibinfo {author} {\bibfnamefont {Z.}~\bibnamefont {Zhao}},\ }\href {https://doi.org/10.1103/PhysRevB.97.224510} {\bibfield  {journal} {\bibinfo  {journal} {Phys. Rev. B}\ }\textbf {\bibinfo {volume} {97}},\ \bibinfo {pages} {224510} (\bibinfo {year} {2018})}\BibitemShut {NoStop}%
\bibitem [{\citenamefont {Friedemann}\ \emph {et~al.}(2018)\citenamefont {Friedemann}, \citenamefont {Duncan}, \citenamefont {Hirschberger}, \citenamefont {Bauer}, \citenamefont {K\"{u}chler}, \citenamefont {Neubauer}, \citenamefont {Brando}, \citenamefont {Pfleiderer},\ and\ \citenamefont {Grosche}}]{QTPex}%
  \BibitemOpen
  \bibfield  {author} {\bibinfo {author} {\bibfnamefont {S.}~\bibnamefont {Friedemann}}, \bibinfo {author} {\bibfnamefont {W.}~\bibnamefont {Duncan}}, \bibinfo {author} {\bibfnamefont {M.}~\bibnamefont {Hirschberger}}, \bibinfo {author} {\bibfnamefont {T.}~\bibnamefont {Bauer}}, \bibinfo {author} {\bibfnamefont {R.}~\bibnamefont {K\"{u}chler}}, \bibinfo {author} {\bibfnamefont {A.}~\bibnamefont {Neubauer}}, \bibinfo {author} {\bibfnamefont {M.}~\bibnamefont {Brando}}, \bibinfo {author} {\bibfnamefont {C.}~\bibnamefont {Pfleiderer}},\ and\ \bibinfo {author} {\bibfnamefont {F.}~\bibnamefont {Grosche}},\ }\href {https://doi.org/10.1038/nphys4242} {\bibfield  {journal} {\bibinfo  {journal} {Nat. Phys.}\ }\textbf {\bibinfo {volume} {14}},\ \bibinfo {pages} {62} (\bibinfo {year} {2018})}\BibitemShut {NoStop}%
\bibitem [{\citenamefont {Nagy}\ \emph {et~al.}(2011)\citenamefont {Nagy}, \citenamefont {Szirmai},\ and\ \citenamefont {Domokos}}]{nagy2011}%
  \BibitemOpen
  \bibfield  {author} {\bibinfo {author} {\bibfnamefont {D.}~\bibnamefont {Nagy}}, \bibinfo {author} {\bibfnamefont {G.}~\bibnamefont {Szirmai}},\ and\ \bibinfo {author} {\bibfnamefont {P.}~\bibnamefont {Domokos}},\ }\href {https://doi.org/10.1103/PhysRevA.84.043637} {\bibfield  {journal} {\bibinfo  {journal} {Phys. Rev. A}\ }\textbf {\bibinfo {volume} {84}},\ \bibinfo {pages} {043637} (\bibinfo {year} {2011})}\BibitemShut {NoStop}%
\bibitem [{\citenamefont {Bhaseen}\ \emph {et~al.}(2012)\citenamefont {Bhaseen}, \citenamefont {Mayoh}, \citenamefont {Simons},\ and\ \citenamefont {Keeling}}]{Bhaasen}%
  \BibitemOpen
  \bibfield  {author} {\bibinfo {author} {\bibfnamefont {M.~J.}\ \bibnamefont {Bhaseen}}, \bibinfo {author} {\bibfnamefont {J.}~\bibnamefont {Mayoh}}, \bibinfo {author} {\bibfnamefont {B.~D.}\ \bibnamefont {Simons}},\ and\ \bibinfo {author} {\bibfnamefont {J.}~\bibnamefont {Keeling}},\ }\href {https://doi.org/10.1103/PhysRevA.85.013817} {\bibfield  {journal} {\bibinfo  {journal} {Phys. Rev. A}\ }\textbf {\bibinfo {volume} {85}},\ \bibinfo {pages} {013817} (\bibinfo {year} {2012})}\BibitemShut {NoStop}%
\bibitem [{\citenamefont {Gelhausen}\ and\ \citenamefont {Buchhold}(2018)}]{gelhausen}%
  \BibitemOpen
  \bibfield  {author} {\bibinfo {author} {\bibfnamefont {J.}~\bibnamefont {Gelhausen}}\ and\ \bibinfo {author} {\bibfnamefont {M.}~\bibnamefont {Buchhold}},\ }\href {https://doi.org/10.1103/PhysRevA.97.023807} {\bibfield  {journal} {\bibinfo  {journal} {Phys. Rev. A}\ }\textbf {\bibinfo {volume} {97}},\ \bibinfo {pages} {023807} (\bibinfo {year} {2018})}\BibitemShut {NoStop}%
\bibitem [{\citenamefont {Soriente}\ \emph {et~al.}(2018)\citenamefont {Soriente}, \citenamefont {Donner}, \citenamefont {Chitra},\ and\ \citenamefont {Zilberberg}}]{Soriente}%
  \BibitemOpen
  \bibfield  {author} {\bibinfo {author} {\bibfnamefont {M.}~\bibnamefont {Soriente}}, \bibinfo {author} {\bibfnamefont {T.}~\bibnamefont {Donner}}, \bibinfo {author} {\bibfnamefont {R.}~\bibnamefont {Chitra}},\ and\ \bibinfo {author} {\bibfnamefont {O.}~\bibnamefont {Zilberberg}},\ }\href {https://doi.org/10.1103/PhysRevLett.120.183603} {\bibfield  {journal} {\bibinfo  {journal} {Phys. Rev. Lett.}\ }\textbf {\bibinfo {volume} {120}},\ \bibinfo {pages} {183603} (\bibinfo {year} {2018})}\BibitemShut {NoStop}%
\bibitem [{\citenamefont {Gegg}\ \emph {et~al.}(2018)\citenamefont {Gegg}, \citenamefont {Carmele}, \citenamefont {Knorr},\ and\ \citenamefont {Richter}}]{Gegg_2018}%
  \BibitemOpen
  \bibfield  {author} {\bibinfo {author} {\bibfnamefont {M.}~\bibnamefont {Gegg}}, \bibinfo {author} {\bibfnamefont {A.}~\bibnamefont {Carmele}}, \bibinfo {author} {\bibfnamefont {A.}~\bibnamefont {Knorr}},\ and\ \bibinfo {author} {\bibfnamefont {M.}~\bibnamefont {Richter}},\ }\href {https://doi.org/10.1088/1367-2630/aa9cdd} {\bibfield  {journal} {\bibinfo  {journal} {New J. Phys.}\ }\textbf {\bibinfo {volume} {20}},\ \bibinfo {pages} {013006} (\bibinfo {year} {2018})}\BibitemShut {NoStop}%
\bibitem [{\citenamefont {{de Aguiar}}\ \emph {et~al.}(1992)\citenamefont {{de Aguiar}}, \citenamefont {Furuya}, \citenamefont {Lewenkopf},\ and\ \citenamefont {Nemes}}]{DEAGUIAR1992291}%
  \BibitemOpen
  \bibfield  {author} {\bibinfo {author} {\bibfnamefont {M.}~\bibnamefont {{de Aguiar}}}, \bibinfo {author} {\bibfnamefont {K.}~\bibnamefont {Furuya}}, \bibinfo {author} {\bibfnamefont {C.}~\bibnamefont {Lewenkopf}},\ and\ \bibinfo {author} {\bibfnamefont {M.}~\bibnamefont {Nemes}},\ }\href {https://doi.org/https://doi.org/10.1016/0003-4916(92)90178-O} {\bibfield  {journal} {\bibinfo  {journal} {Annals of Physics}\ }\textbf {\bibinfo {volume} {216}},\ \bibinfo {pages} {291} (\bibinfo {year} {1992})}\BibitemShut {NoStop}%
\bibitem [{\citenamefont {Furuya}\ \emph {et~al.}(1992)\citenamefont {Furuya}, \citenamefont {{de Aguiar}}, \citenamefont {Lewenkopf},\ and\ \citenamefont {Nemes}}]{FURUYA1992313}%
  \BibitemOpen
  \bibfield  {author} {\bibinfo {author} {\bibfnamefont {K.}~\bibnamefont {Furuya}}, \bibinfo {author} {\bibfnamefont {M.}~\bibnamefont {{de Aguiar}}}, \bibinfo {author} {\bibfnamefont {C.}~\bibnamefont {Lewenkopf}},\ and\ \bibinfo {author} {\bibfnamefont {M.}~\bibnamefont {Nemes}},\ }\href {https://doi.org/https://doi.org/10.1016/0003-4916(92)90179-P} {\bibfield  {journal} {\bibinfo  {journal} {Annals of Physics}\ }\textbf {\bibinfo {volume} {216}},\ \bibinfo {pages} {313} (\bibinfo {year} {1992})}\BibitemShut {NoStop}%
\bibitem [{\citenamefont {Bastarrachea-Magnani}\ \emph {et~al.}(2016)\citenamefont {Bastarrachea-Magnani}, \citenamefont {Lerma-Hernández},\ and\ \citenamefont {Hirsch}}]{Bastarrachea-Magnani_2016}%
  \BibitemOpen
  \bibfield  {author} {\bibinfo {author} {\bibfnamefont {M.~A.}\ \bibnamefont {Bastarrachea-Magnani}}, \bibinfo {author} {\bibfnamefont {S.}~\bibnamefont {Lerma-Hernández}},\ and\ \bibinfo {author} {\bibfnamefont {J.~G.}\ \bibnamefont {Hirsch}},\ }\href {https://doi.org/10.1088/1742-5468/2016/09/093105} {\bibfield  {journal} {\bibinfo  {journal} {Journal of Statistical Mechanics: Theory and Experiment}\ }\textbf {\bibinfo {volume} {2016}},\ \bibinfo {pages} {093105} (\bibinfo {year} {2016})}\BibitemShut {NoStop}%
\bibitem [{\citenamefont {Kloc}\ \emph {et~al.}(2017)\citenamefont {Kloc}, \citenamefont {Stránský},\ and\ \citenamefont {Cejnar}}]{KLOC201785}%
  \BibitemOpen
  \bibfield  {author} {\bibinfo {author} {\bibfnamefont {M.}~\bibnamefont {Kloc}}, \bibinfo {author} {\bibfnamefont {P.}~\bibnamefont {Stránský}},\ and\ \bibinfo {author} {\bibfnamefont {P.}~\bibnamefont {Cejnar}},\ }\href {https://doi.org/https://doi.org/10.1016/j.aop.2017.04.005} {\bibfield  {journal} {\bibinfo  {journal} {Annals of Physics}\ }\textbf {\bibinfo {volume} {382}},\ \bibinfo {pages} {85} (\bibinfo {year} {2017})}\BibitemShut {NoStop}%
\bibitem [{\citenamefont {Baksic}\ and\ \citenamefont {Ciuti}(2014)}]{Baksic}%
  \BibitemOpen
  \bibfield  {author} {\bibinfo {author} {\bibfnamefont {A.}~\bibnamefont {Baksic}}\ and\ \bibinfo {author} {\bibfnamefont {C.}~\bibnamefont {Ciuti}},\ }\href {https://doi.org/10.1103/PhysRevLett.112.173601} {\bibfield  {journal} {\bibinfo  {journal} {Phys. Rev. Lett.}\ }\textbf {\bibinfo {volume} {112}},\ \bibinfo {pages} {173601} (\bibinfo {year} {2014})}\BibitemShut {NoStop}%
\bibitem [{\citenamefont {Cordero}\ \emph {et~al.}(2013)\citenamefont {Cordero}, \citenamefont {L\'opez-Pe\~na}, \citenamefont {Casta\~nos},\ and\ \citenamefont {Nahmad-Achar}}]{cordero2013}%
  \BibitemOpen
  \bibfield  {author} {\bibinfo {author} {\bibfnamefont {S.}~\bibnamefont {Cordero}}, \bibinfo {author} {\bibfnamefont {R.}~\bibnamefont {L\'opez-Pe\~na}}, \bibinfo {author} {\bibfnamefont {O.}~\bibnamefont {Casta\~nos}},\ and\ \bibinfo {author} {\bibfnamefont {E.}~\bibnamefont {Nahmad-Achar}},\ }\href {https://doi.org/10.1103/PhysRevA.87.023805} {\bibfield  {journal} {\bibinfo  {journal} {Phys. Rev. A}\ }\textbf {\bibinfo {volume} {87}},\ \bibinfo {pages} {023805} (\bibinfo {year} {2013})}\BibitemShut {NoStop}%
\bibitem [{\citenamefont {Rza\ifmmode~\dot{z}\else \.{z}\fi{}ewski}\ \emph {et~al.}(1975)\citenamefont {Rza\ifmmode~\dot{z}\else \.{z}\fi{}ewski}, \citenamefont {W\'odkiewicz},\ and\ \citenamefont {\ifmmode~\dot{Z}\else \.{Z}\fi{}akowicz}}]{rzazewski}%
  \BibitemOpen
  \bibfield  {author} {\bibinfo {author} {\bibfnamefont {K.}~\bibnamefont {Rza\ifmmode~\dot{z}\else \.{z}\fi{}ewski}}, \bibinfo {author} {\bibfnamefont {K.}~\bibnamefont {W\'odkiewicz}},\ and\ \bibinfo {author} {\bibfnamefont {W.}~\bibnamefont {\ifmmode~\dot{Z}\else \.{Z}\fi{}akowicz}},\ }\href {https://doi.org/10.1103/PhysRevLett.35.432} {\bibfield  {journal} {\bibinfo  {journal} {Phys. Rev. Lett.}\ }\textbf {\bibinfo {volume} {35}},\ \bibinfo {pages} {432} (\bibinfo {year} {1975})}\BibitemShut {NoStop}%
\bibitem [{\citenamefont {Dimer}\ \emph {et~al.}(2007)\citenamefont {Dimer}, \citenamefont {Estienne}, \citenamefont {Parkins},\ and\ \citenamefont {Carmichael}}]{Dimer}%
  \BibitemOpen
  \bibfield  {author} {\bibinfo {author} {\bibfnamefont {F.}~\bibnamefont {Dimer}}, \bibinfo {author} {\bibfnamefont {B.}~\bibnamefont {Estienne}}, \bibinfo {author} {\bibfnamefont {A.~S.}\ \bibnamefont {Parkins}},\ and\ \bibinfo {author} {\bibfnamefont {H.~J.}\ \bibnamefont {Carmichael}},\ }\href {https://doi.org/10.1103/PhysRevA.75.013804} {\bibfield  {journal} {\bibinfo  {journal} {Phys. Rev. A}\ }\textbf {\bibinfo {volume} {75}},\ \bibinfo {pages} {013804} (\bibinfo {year} {2007})}\BibitemShut {NoStop}%
\bibitem [{\citenamefont {Léonard}\ \emph {et~al.}(2017)\citenamefont {Léonard}, \citenamefont {Morales}, \citenamefont {Zupancic}, \citenamefont {Donner},\ and\ \citenamefont {Esslinger}}]{leonard}%
  \BibitemOpen
  \bibfield  {author} {\bibinfo {author} {\bibfnamefont {J.}~\bibnamefont {Léonard}}, \bibinfo {author} {\bibfnamefont {A.}~\bibnamefont {Morales}}, \bibinfo {author} {\bibfnamefont {P.}~\bibnamefont {Zupancic}}, \bibinfo {author} {\bibfnamefont {T.}~\bibnamefont {Donner}},\ and\ \bibinfo {author} {\bibfnamefont {T.}~\bibnamefont {Esslinger}},\ }\href {https://doi.org/10.1126/science.aan2608} {\bibfield  {journal} {\bibinfo  {journal} {Science}\ }\textbf {\bibinfo {volume} {358}},\ \bibinfo {pages} {1415} (\bibinfo {year} {2017})},\ \Eprint {https://arxiv.org/abs/https://www.science.org/doi/pdf/10.1126/science.aan2608} {https://www.science.org/doi/pdf/10.1126/science.aan2608} \BibitemShut {NoStop}%
\bibitem [{\citenamefont {López-Peña}\ \emph {et~al.}(2021)\citenamefont {López-Peña}, \citenamefont {Cordero}, \citenamefont {Nahmad-Achar},\ and\ \citenamefont {Castaños}}]{Lopez2021}%
  \BibitemOpen
  \bibfield  {author} {\bibinfo {author} {\bibfnamefont {R.}~\bibnamefont {López-Peña}}, \bibinfo {author} {\bibfnamefont {S.}~\bibnamefont {Cordero}}, \bibinfo {author} {\bibfnamefont {E.}~\bibnamefont {Nahmad-Achar}},\ and\ \bibinfo {author} {\bibfnamefont {O.}~\bibnamefont {Castaños}},\ }\href {https://doi.org/10.1088/1402-4896/abd654} {\bibfield  {journal} {\bibinfo  {journal} {Physica Scripta}\ }\textbf {\bibinfo {volume} {96}},\ \bibinfo {pages} {035103} (\bibinfo {year} {2021})}\BibitemShut {NoStop}%
\bibitem [{\citenamefont {Cordero}\ \emph {et~al.}(2021)\citenamefont {Cordero}, \citenamefont {Nahmad-Achar}, \citenamefont {López-Peña},\ and\ \citenamefont {Castaños}}]{Cordero_2021}%
  \BibitemOpen
  \bibfield  {author} {\bibinfo {author} {\bibfnamefont {S.}~\bibnamefont {Cordero}}, \bibinfo {author} {\bibfnamefont {E.}~\bibnamefont {Nahmad-Achar}}, \bibinfo {author} {\bibfnamefont {R.}~\bibnamefont {López-Peña}},\ and\ \bibinfo {author} {\bibfnamefont {O.}~\bibnamefont {Castaños}},\ }\href {https://doi.org/10.1088/1402-4896/abd653} {\bibfield  {journal} {\bibinfo  {journal} {Physica Scripta}\ }\textbf {\bibinfo {volume} {96}},\ \bibinfo {pages} {035104} (\bibinfo {year} {2021})}\BibitemShut {NoStop}%
\bibitem [{\citenamefont {Larson}\ and\ \citenamefont {Mavrogordatos}(2021)}]{Larson_2021}%
  \BibitemOpen
  \bibfield  {author} {\bibinfo {author} {\bibfnamefont {J.}~\bibnamefont {Larson}}\ and\ \bibinfo {author} {\bibfnamefont {T.}~\bibnamefont {Mavrogordatos}},\ }\href {https://doi.org/10.1088/978-0-7503-3447-1} {\emph {\bibinfo {title} {The Jaynes–Cummings Model and Its Descendants}}},\ 2053-2563\ (\bibinfo  {publisher} {IOP Publishing},\ \bibinfo {year} {2021})\BibitemShut {NoStop}%
\bibitem [{\citenamefont {Holstein}\ and\ \citenamefont {Primakoff}(1940)}]{HP1940}%
  \BibitemOpen
  \bibfield  {author} {\bibinfo {author} {\bibfnamefont {T.}~\bibnamefont {Holstein}}\ and\ \bibinfo {author} {\bibfnamefont {H.}~\bibnamefont {Primakoff}},\ }\href {https://doi.org/10.1103/PhysRev.58.1098} {\bibfield  {journal} {\bibinfo  {journal} {Phys. Rev.}\ }\textbf {\bibinfo {volume} {58}},\ \bibinfo {pages} {1098} (\bibinfo {year} {1940})}\BibitemShut {NoStop}%
\bibitem [{\citenamefont {Kirton}\ \emph {et~al.}(2019)\citenamefont {Kirton}, \citenamefont {Roses}, \citenamefont {Keeling},\ and\ \citenamefont {Dalla~Torre}}]{kirton2019}%
  \BibitemOpen
  \bibfield  {author} {\bibinfo {author} {\bibfnamefont {P.}~\bibnamefont {Kirton}}, \bibinfo {author} {\bibfnamefont {M.~M.}\ \bibnamefont {Roses}}, \bibinfo {author} {\bibfnamefont {J.}~\bibnamefont {Keeling}},\ and\ \bibinfo {author} {\bibfnamefont {E.~G.}\ \bibnamefont {Dalla~Torre}},\ }\href {https://doi.org/https://doi.org/10.1002/qute.201800043} {\bibfield  {journal} {\bibinfo  {journal} {Adv. Quantum Tech.}\ }\textbf {\bibinfo {volume} {2}},\ \bibinfo {pages} {1800043} (\bibinfo {year} {2019})}\BibitemShut {NoStop}%
\bibitem [{\citenamefont {Bogoljubov}(1958)}]{Bogoljubov1958-te}%
  \BibitemOpen
  \bibfield  {author} {\bibinfo {author} {\bibfnamefont {N.~N.}\ \bibnamefont {Bogoljubov}},\ }\href {https://doi.org/10.1007/BF02745585} {\bibfield  {journal} {\bibinfo  {journal} {Il Nuovo Cimento}\ }\textbf {\bibinfo {volume} {7}},\ \bibinfo {pages} {794} (\bibinfo {year} {1958})}\BibitemShut {NoStop}%
\bibitem [{\citenamefont {Valatin}(1958)}]{valatin1958comments}%
  \BibitemOpen
  \bibfield  {author} {\bibinfo {author} {\bibfnamefont {J.}~\bibnamefont {Valatin}},\ }\href {https://doi.org/10.1007/BF02745589} {\bibfield  {journal} {\bibinfo  {journal} {Il Nuovo Cimento (1955-1965)}\ }\textbf {\bibinfo {volume} {7}},\ \bibinfo {pages} {843} (\bibinfo {year} {1958})}\BibitemShut {NoStop}%
\bibitem [{\citenamefont {Hirsch}\ \emph {et~al.}(2013)\citenamefont {Hirsch}, \citenamefont {Castaños}, \citenamefont {López-Peña},\ and\ \citenamefont {Nahmad-Achar}}]{Hirsch_2013}%
  \BibitemOpen
  \bibfield  {author} {\bibinfo {author} {\bibfnamefont {J.~G.}\ \bibnamefont {Hirsch}}, \bibinfo {author} {\bibfnamefont {O.}~\bibnamefont {Castaños}}, \bibinfo {author} {\bibfnamefont {R.}~\bibnamefont {López-Peña}},\ and\ \bibinfo {author} {\bibfnamefont {E.}~\bibnamefont {Nahmad-Achar}},\ }\href {https://doi.org/10.1088/0031-8949/87/03/038106} {\bibfield  {journal} {\bibinfo  {journal} {Physica Scripta}\ }\textbf {\bibinfo {volume} {87}},\ \bibinfo {pages} {038106} (\bibinfo {year} {2013})}\BibitemShut {NoStop}%
\bibitem [{\citenamefont {Goldstone}\ \emph {et~al.}(1962)\citenamefont {Goldstone}, \citenamefont {Salam},\ and\ \citenamefont {Weinberg}}]{goldstone}%
  \BibitemOpen
  \bibfield  {author} {\bibinfo {author} {\bibfnamefont {J.}~\bibnamefont {Goldstone}}, \bibinfo {author} {\bibfnamefont {A.}~\bibnamefont {Salam}},\ and\ \bibinfo {author} {\bibfnamefont {S.}~\bibnamefont {Weinberg}},\ }\href {https://doi.org/10.1103/PhysRev.127.965} {\bibfield  {journal} {\bibinfo  {journal} {Phys. Rev.}\ }\textbf {\bibinfo {volume} {127}},\ \bibinfo {pages} {965} (\bibinfo {year} {1962})}\BibitemShut {NoStop}%
\bibitem [{\citenamefont {Gell-Mann}(1962)}]{gellmann}%
  \BibitemOpen
  \bibfield  {author} {\bibinfo {author} {\bibfnamefont {M.}~\bibnamefont {Gell-Mann}},\ }\href {https://doi.org/10.1103/PhysRev.125.1067} {\bibfield  {journal} {\bibinfo  {journal} {Phys. Rev.}\ }\textbf {\bibinfo {volume} {125}},\ \bibinfo {pages} {1067} (\bibinfo {year} {1962})}\BibitemShut {NoStop}%
\bibitem [{\citenamefont {PAIS}(1966)}]{pais}%
  \BibitemOpen
  \bibfield  {author} {\bibinfo {author} {\bibfnamefont {A.}~\bibnamefont {PAIS}},\ }\href {https://doi.org/10.1103/RevModPhys.38.215} {\bibfield  {journal} {\bibinfo  {journal} {Rev. Mod. Phys.}\ }\textbf {\bibinfo {volume} {38}},\ \bibinfo {pages} {215} (\bibinfo {year} {1966})}\BibitemShut {NoStop}%
\bibitem [{\citenamefont {Greiner}\ and\ \citenamefont {M{\"u}ller}(2012)}]{greiner2012quantum}%
  \BibitemOpen
  \bibfield  {author} {\bibinfo {author} {\bibfnamefont {W.}~\bibnamefont {Greiner}}\ and\ \bibinfo {author} {\bibfnamefont {B.}~\bibnamefont {M{\"u}ller}},\ }\href@noop {} {\emph {\bibinfo {title} {Quantum mechanics: symmetries}}}\ (\bibinfo  {publisher} {Springer Science \& Business Media},\ \bibinfo {year} {2012})\BibitemShut {NoStop}%
\bibitem [{\citenamefont {Baird}\ and\ \citenamefont {Biedenharn}(1963)}]{baird1963}%
  \BibitemOpen
  \bibfield  {author} {\bibinfo {author} {\bibfnamefont {G.~E.}\ \bibnamefont {Baird}}\ and\ \bibinfo {author} {\bibfnamefont {L.~C.}\ \bibnamefont {Biedenharn}},\ }\href {https://doi.org/10.1063/1.1703926} {\bibfield  {journal} {\bibinfo  {journal} {Journal of Mathematical Physics}\ }\textbf {\bibinfo {volume} {4}},\ \bibinfo {pages} {1449} (\bibinfo {year} {1963})}\BibitemShut {NoStop}%
\bibitem [{\citenamefont {Macfarlane}\ \emph {et~al.}(1963)\citenamefont {Macfarlane}, \citenamefont {Sudarshan},\ and\ \citenamefont {Dullemond}}]{macfarlane1963weyl}%
  \BibitemOpen
  \bibfield  {author} {\bibinfo {author} {\bibfnamefont {A.}~\bibnamefont {Macfarlane}}, \bibinfo {author} {\bibfnamefont {E.}~\bibnamefont {Sudarshan}},\ and\ \bibinfo {author} {\bibfnamefont {C.}~\bibnamefont {Dullemond}},\ }\href {https://doi.org/10.1007/BF02750419} {\bibfield  {journal} {\bibinfo  {journal} {Il Nuovo Cimento (1955-1965)}\ }\textbf {\bibinfo {volume} {30}},\ \bibinfo {pages} {845} (\bibinfo {year} {1963})}\BibitemShut {NoStop}%
\bibitem [{\citenamefont {Rosso}\ \emph {et~al.}(2022)\citenamefont {Rosso}, \citenamefont {Mazza},\ and\ \citenamefont {Biella}}]{Rosso2022}%
  \BibitemOpen
  \bibfield  {author} {\bibinfo {author} {\bibfnamefont {L.}~\bibnamefont {Rosso}}, \bibinfo {author} {\bibfnamefont {L.}~\bibnamefont {Mazza}},\ and\ \bibinfo {author} {\bibfnamefont {A.}~\bibnamefont {Biella}},\ }\href {https://doi.org/10.1103/PhysRevA.105.L051302} {\bibfield  {journal} {\bibinfo  {journal} {Phys. Rev. A}\ }\textbf {\bibinfo {volume} {105}},\ \bibinfo {pages} {L051302} (\bibinfo {year} {2022})}\BibitemShut {NoStop}%
\bibitem [{\citenamefont {Hartmann}(2016)}]{hartmann}%
  \BibitemOpen
  \bibfield  {author} {\bibinfo {author} {\bibfnamefont {S.}~\bibnamefont {Hartmann}},\ }\href@noop {} {\bibfield  {journal} {\bibinfo  {journal} {Quantum Inf. Comput.}\ }\textbf {\bibinfo {volume} {16}},\ \bibinfo {pages} {1333} (\bibinfo {year} {2016})}\BibitemShut {NoStop}%
\bibitem [{\citenamefont {Masson}\ \emph {et~al.}(2017)\citenamefont {Masson}, \citenamefont {Barrett},\ and\ \citenamefont {Parkins}}]{Masson2017}%
  \BibitemOpen
  \bibfield  {author} {\bibinfo {author} {\bibfnamefont {S.~J.}\ \bibnamefont {Masson}}, \bibinfo {author} {\bibfnamefont {M.~D.}\ \bibnamefont {Barrett}},\ and\ \bibinfo {author} {\bibfnamefont {S.}~\bibnamefont {Parkins}},\ }\href {https://doi.org/10.1103/PhysRevLett.119.213601} {\bibfield  {journal} {\bibinfo  {journal} {Phys. Rev. Lett.}\ }\textbf {\bibinfo {volume} {119}},\ \bibinfo {pages} {213601} (\bibinfo {year} {2017})}\BibitemShut {NoStop}%
\bibitem [{\citenamefont {Zhiqiang}\ \emph {et~al.}(2017)\citenamefont {Zhiqiang}, \citenamefont {Lee}, \citenamefont {Kumar}, \citenamefont {Arnold}, \citenamefont {Masson}, \citenamefont {Parkins},\ and\ \citenamefont {Barrett}}]{Zhiqiang:17}%
  \BibitemOpen
  \bibfield  {author} {\bibinfo {author} {\bibfnamefont {Z.}~\bibnamefont {Zhiqiang}}, \bibinfo {author} {\bibfnamefont {C.~H.}\ \bibnamefont {Lee}}, \bibinfo {author} {\bibfnamefont {R.}~\bibnamefont {Kumar}}, \bibinfo {author} {\bibfnamefont {K.~J.}\ \bibnamefont {Arnold}}, \bibinfo {author} {\bibfnamefont {S.~J.}\ \bibnamefont {Masson}}, \bibinfo {author} {\bibfnamefont {A.~S.}\ \bibnamefont {Parkins}},\ and\ \bibinfo {author} {\bibfnamefont {M.~D.}\ \bibnamefont {Barrett}},\ }\href {https://doi.org/10.1364/OPTICA.4.000424} {\bibfield  {journal} {\bibinfo  {journal} {Optica}\ }\textbf {\bibinfo {volume} {4}},\ \bibinfo {pages} {424} (\bibinfo {year} {2017})}\BibitemShut {NoStop}%
\bibitem [{\citenamefont {Overbeck}\ \emph {et~al.}(2017)\citenamefont {Overbeck}, \citenamefont {Maghrebi}, \citenamefont {Gorshkov},\ and\ \citenamefont {Weimer}}]{triIsing1}%
  \BibitemOpen
  \bibfield  {author} {\bibinfo {author} {\bibfnamefont {V.~R.}\ \bibnamefont {Overbeck}}, \bibinfo {author} {\bibfnamefont {M.~F.}\ \bibnamefont {Maghrebi}}, \bibinfo {author} {\bibfnamefont {A.~V.}\ \bibnamefont {Gorshkov}},\ and\ \bibinfo {author} {\bibfnamefont {H.}~\bibnamefont {Weimer}},\ }\href {https://doi.org/10.1103/PhysRevA.95.042133} {\bibfield  {journal} {\bibinfo  {journal} {Phys. Rev. A}\ }\textbf {\bibinfo {volume} {95}},\ \bibinfo {pages} {042133} (\bibinfo {year} {2017})}\BibitemShut {NoStop}%
\bibitem [{\citenamefont {Baumann}\ \emph {et~al.}(2010)\citenamefont {Baumann}, \citenamefont {Guerlin}, \citenamefont {Brennecke},\ and\ \citenamefont {Esslinger}}]{Baumann2010-zm}%
  \BibitemOpen
  \bibfield  {author} {\bibinfo {author} {\bibfnamefont {K.}~\bibnamefont {Baumann}}, \bibinfo {author} {\bibfnamefont {C.}~\bibnamefont {Guerlin}}, \bibinfo {author} {\bibfnamefont {F.}~\bibnamefont {Brennecke}},\ and\ \bibinfo {author} {\bibfnamefont {T.}~\bibnamefont {Esslinger}},\ }\href {https://doi.org/10.1038/nature09009} {\bibfield  {journal} {\bibinfo  {journal} {Nature}\ }\textbf {\bibinfo {volume} {464}},\ \bibinfo {pages} {1301} (\bibinfo {year} {2010})}\BibitemShut {NoStop}%
\bibitem [{\citenamefont {Li}\ \emph {et~al.}(2018)\citenamefont {Li}, \citenamefont {Bamba}, \citenamefont {Yuan}, \citenamefont {Zhang}, \citenamefont {Zhao}, \citenamefont {Xiang}, \citenamefont {Xu}, \citenamefont {Jin}, \citenamefont {Ren}, \citenamefont {Ma}, \citenamefont {Cao}, \citenamefont {Turchinovich},\ and\ \citenamefont {Kono}}]{li2018}%
  \BibitemOpen
  \bibfield  {author} {\bibinfo {author} {\bibfnamefont {X.}~\bibnamefont {Li}}, \bibinfo {author} {\bibfnamefont {M.}~\bibnamefont {Bamba}}, \bibinfo {author} {\bibfnamefont {N.}~\bibnamefont {Yuan}}, \bibinfo {author} {\bibfnamefont {Q.}~\bibnamefont {Zhang}}, \bibinfo {author} {\bibfnamefont {Y.}~\bibnamefont {Zhao}}, \bibinfo {author} {\bibfnamefont {M.}~\bibnamefont {Xiang}}, \bibinfo {author} {\bibfnamefont {K.}~\bibnamefont {Xu}}, \bibinfo {author} {\bibfnamefont {Z.}~\bibnamefont {Jin}}, \bibinfo {author} {\bibfnamefont {W.}~\bibnamefont {Ren}}, \bibinfo {author} {\bibfnamefont {G.}~\bibnamefont {Ma}}, \bibinfo {author} {\bibfnamefont {S.}~\bibnamefont {Cao}}, \bibinfo {author} {\bibfnamefont {D.}~\bibnamefont {Turchinovich}},\ and\ \bibinfo {author} {\bibfnamefont {J.}~\bibnamefont {Kono}},\ }\href {https://doi.org/10.1126/science.aat5162} {\bibfield  {journal} {\bibinfo  {journal} {Science}\ }\textbf {\bibinfo {volume} {361}},\ \bibinfo {pages} {794} (\bibinfo {year} {2018})},\ \Eprint
  {https://arxiv.org/abs/https://www.science.org/doi/pdf/10.1126/science.aat5162} {https://www.science.org/doi/pdf/10.1126/science.aat5162} \BibitemShut {NoStop}%
\bibitem [{\citenamefont {Mezzacapo}\ \emph {et~al.}(2014)\citenamefont {Mezzacapo}, \citenamefont {Las~Heras}, \citenamefont {Pedernales}, \citenamefont {DiCarlo}, \citenamefont {Solano},\ and\ \citenamefont {Lamata}}]{Mezzacapo2014-xl}%
  \BibitemOpen
  \bibfield  {author} {\bibinfo {author} {\bibfnamefont {A.}~\bibnamefont {Mezzacapo}}, \bibinfo {author} {\bibfnamefont {U.}~\bibnamefont {Las~Heras}}, \bibinfo {author} {\bibfnamefont {J.~S.}\ \bibnamefont {Pedernales}}, \bibinfo {author} {\bibfnamefont {L.}~\bibnamefont {DiCarlo}}, \bibinfo {author} {\bibfnamefont {E.}~\bibnamefont {Solano}},\ and\ \bibinfo {author} {\bibfnamefont {L.}~\bibnamefont {Lamata}},\ }\href {https://doi.org/10.1038/srep07482} {\bibfield  {journal} {\bibinfo  {journal} {Sci. Rep.}\ }\textbf {\bibinfo {volume} {4}},\ \bibinfo {pages} {7482} (\bibinfo {year} {2014})}\BibitemShut {NoStop}%
\bibitem [{\citenamefont {Campbell}\ \emph {et~al.}(2016)\citenamefont {Campbell}, \citenamefont {Price}, \citenamefont {Putra}, \citenamefont {Vald{\'e}s-Curiel}, \citenamefont {Trypogeorgos},\ and\ \citenamefont {Spielman}}]{campbell2016magnetic}%
  \BibitemOpen
  \bibfield  {author} {\bibinfo {author} {\bibfnamefont {D.}~\bibnamefont {Campbell}}, \bibinfo {author} {\bibfnamefont {R.}~\bibnamefont {Price}}, \bibinfo {author} {\bibfnamefont {A.}~\bibnamefont {Putra}}, \bibinfo {author} {\bibfnamefont {A.}~\bibnamefont {Vald{\'e}s-Curiel}}, \bibinfo {author} {\bibfnamefont {D.}~\bibnamefont {Trypogeorgos}},\ and\ \bibinfo {author} {\bibfnamefont {I.}~\bibnamefont {Spielman}},\ }\href {https://doi.org/10.1038/ncomms10897} {\bibfield  {journal} {\bibinfo  {journal} {Nat. Commun.}\ }\textbf {\bibinfo {volume} {7}},\ \bibinfo {pages} {10897} (\bibinfo {year} {2016})}\BibitemShut {NoStop}%
\bibitem [{\citenamefont {Hamner}\ \emph {et~al.}(2014)\citenamefont {Hamner}, \citenamefont {Qu}, \citenamefont {Zhang}, \citenamefont {Chang}, \citenamefont {Gong}, \citenamefont {Zhang},\ and\ \citenamefont {Engels}}]{dickesoc12345}%
  \BibitemOpen
  \bibfield  {author} {\bibinfo {author} {\bibfnamefont {C.}~\bibnamefont {Hamner}}, \bibinfo {author} {\bibfnamefont {C.}~\bibnamefont {Qu}}, \bibinfo {author} {\bibfnamefont {Y.}~\bibnamefont {Zhang}}, \bibinfo {author} {\bibfnamefont {J.}~\bibnamefont {Chang}}, \bibinfo {author} {\bibfnamefont {M.}~\bibnamefont {Gong}}, \bibinfo {author} {\bibfnamefont {C.}~\bibnamefont {Zhang}},\ and\ \bibinfo {author} {\bibfnamefont {P.}~\bibnamefont {Engels}},\ }\href {https://doi.org/10.1038/ncomms5023} {\bibfield  {journal} {\bibinfo  {journal} {Nat. Commun.}\ }\textbf {\bibinfo {volume} {5}},\ \bibinfo {pages} {4023} (\bibinfo {year} {2014})}\BibitemShut {NoStop}%
\bibitem [{\citenamefont {Chaikin}\ and\ \citenamefont {Lubensky}(1995)}]{chaikin_lubensky_1995}%
  \BibitemOpen
  \bibfield  {author} {\bibinfo {author} {\bibfnamefont {P.~M.}\ \bibnamefont {Chaikin}}\ and\ \bibinfo {author} {\bibfnamefont {T.~C.}\ \bibnamefont {Lubensky}},\ }\href {https://doi.org/10.1017/CBO9780511813467} {\emph {\bibinfo {title} {Principles of Condensed Matter Physics}}}\ (\bibinfo  {publisher} {Cambridge University Press},\ \bibinfo {year} {1995})\BibitemShut {NoStop}%
\bibitem [{\citenamefont {de~Swart}(1963)}]{deSwart}%
  \BibitemOpen
  \bibfield  {author} {\bibinfo {author} {\bibfnamefont {J.~J.}\ \bibnamefont {de~Swart}},\ }\href {https://doi.org/10.1103/RevModPhys.35.916} {\bibfield  {journal} {\bibinfo  {journal} {Rev. Mod. Phys.}\ }\textbf {\bibinfo {volume} {35}},\ \bibinfo {pages} {916} (\bibinfo {year} {1963})}\BibitemShut {NoStop}%
\bibitem [{\citenamefont {Gerbier}\ \emph {et~al.}(2006)\citenamefont {Gerbier}, \citenamefont {Widera}, \citenamefont {F\"olling}, \citenamefont {Mandel},\ and\ \citenamefont {Bloch}}]{PhysRevA.73.041602}%
  \BibitemOpen
  \bibfield  {author} {\bibinfo {author} {\bibfnamefont {F.}~\bibnamefont {Gerbier}}, \bibinfo {author} {\bibfnamefont {A.}~\bibnamefont {Widera}}, \bibinfo {author} {\bibfnamefont {S.}~\bibnamefont {F\"olling}}, \bibinfo {author} {\bibfnamefont {O.}~\bibnamefont {Mandel}},\ and\ \bibinfo {author} {\bibfnamefont {I.}~\bibnamefont {Bloch}},\ }\href {https://doi.org/10.1103/PhysRevA.73.041602} {\bibfield  {journal} {\bibinfo  {journal} {Phys. Rev. A}\ }\textbf {\bibinfo {volume} {73}},\ \bibinfo {pages} {041602} (\bibinfo {year} {2006})}\BibitemShut {NoStop}%
\bibitem [{\citenamefont {Jensen}\ \emph {et~al.}(2009)\citenamefont {Jensen}, \citenamefont {Acosta}, \citenamefont {Higbie}, \citenamefont {Ledbetter}, \citenamefont {Rochester},\ and\ \citenamefont {Budker}}]{PhysRevA.79.023406}%
  \BibitemOpen
  \bibfield  {author} {\bibinfo {author} {\bibfnamefont {K.}~\bibnamefont {Jensen}}, \bibinfo {author} {\bibfnamefont {V.~M.}\ \bibnamefont {Acosta}}, \bibinfo {author} {\bibfnamefont {J.~M.}\ \bibnamefont {Higbie}}, \bibinfo {author} {\bibfnamefont {M.~P.}\ \bibnamefont {Ledbetter}}, \bibinfo {author} {\bibfnamefont {S.~M.}\ \bibnamefont {Rochester}},\ and\ \bibinfo {author} {\bibfnamefont {D.}~\bibnamefont {Budker}},\ }\href {https://doi.org/10.1103/PhysRevA.79.023406} {\bibfield  {journal} {\bibinfo  {journal} {Phys. Rev. A}\ }\textbf {\bibinfo {volume} {79}},\ \bibinfo {pages} {023406} (\bibinfo {year} {2009})}\BibitemShut {NoStop}%
\end{thebibliography}%

\end{document}